\documentclass[lettersize,journal]{IEEEtran}
\usepackage{cite}
\usepackage{amsmath,amsfonts,amsthm,amssymb}
\usepackage{algorithmic}
\usepackage{array}
\usepackage[caption=false,font=normalsize,labelfont=sf,textfont=sf]{subfig}
\usepackage{textcomp}
\usepackage{stfloats}
\usepackage{url}
\usepackage{verbatim}
\usepackage{graphicx}
\usepackage{xcolor}
\usepackage{multirow}
\usepackage{booktabs}
\usepackage{makecell}
\usepackage[hidelinks]{hyperref}
\usepackage{orcidlink}
\usepackage{bm,bbm}

\DeclareMathOperator*{\argmax}{arg\,max}

\begin{document}
	
\title{Background-Guided Training of Intelligent Radar Target Detectors Under Incomplete Label Supervision}

\author{Chuanfei Zang\orcidlink{0009-0005-8851-5926}, 
	Yiru Lin\orcidlink{0009-0004-6850-5344},
	Yumiao Wang\orcidlink{0000-0002-1474-0486},
	Xingyu Chen\orcidlink{0009-0001-4295-9928},
	and Guolong Cui\orcidlink{0000-0001-5707-6311},~\IEEEmembership{Senior Member,~IEEE}
	\thanks{This work was supported by the National Natural Science Foundation of China under Grant 62271126 and by the Sichuan Province Innovative Talent Funding Project for Postdoctoral Fellows. (\textit{Corresponding author: Yumiao Wang}.)}
	
	\thanks{Chuanfei Zang, Yiru Lin, Yumiao Wang, and Xingyu Chen are with the School of Information and Communication Engineering, University of Electronic Science and Technology of China, Chengdu 62601133, China (e-mail: zangcf2017@163.com; linyiru2021@163.com; wangyumiao@uestc.edu.cn; cxytoby@163.com). }
	
	\thanks{Guolong Cui is with the School of Information and Communication Engineering, University of Electronic Science and Technology of China, Chengdu 611731, China, and also Tianfu Jiangxi Laboratory, Chengdu, Sichuan 641419, China (e-mail: cuiguolong@uestc.edu.cn). }}

\maketitle

\begin{abstract}
	Intelligent radar target detectors based on deep learning can outperform conventional detectors in complex environments, but the range-cell labels typically required for training are difficult to obtain because each target must be localized to a specific range cell. To reduce this annotation burden, we propose a background-guided training method that requires only segment-level labels. These labels provide incomplete supervision for range-cell detection because they indicate whether an echo segment contains a target without specifying the target’s range cell. Firstly, we learn the statistical distribution of the background clutter by fitting a background model to target-free echo segments. For each echo segment containing a target, we use the background model to identify likely target locations and generate refined range-cell pseudo-labels. Secondly, we construct a multilevel loss that combines location supervision from these pseudo-labels with target presence supervision from the incomplete labels, guiding the detector to enhance target responses and suppress background false alarms. Finally, we design a background-aware detector whose input combines the echo signal with whitened residuals and anomaly scores generated by the background model, allowing the detector to make fuller use of background information. In addition, it extracts local features at multiple scales using parallel convolutions and then uses attention to fuse local target features with global background information, further improving its ability to distinguish targets from the background. Experiments on publicly available measured radar echoes show that the proposed method reduces the required signal-to-clutter ratio by more than 13 dB on average relative to our previous method and approaches the performance of a detector trained with accurate range-cell labels.
\end{abstract}


\section{Introduction}
Radar provides long-range sensing and can operate day and night under varied weather conditions, enabling its widespread use in remote sensing, area surveillance, and emergency search and rescue~\cite{11059945,9833364,10098627}. Target detection is one of the key technologies supporting these applications and aims to determine whether a target is present against a background of clutter and noise and to identify its location~\cite{11193878}. Conventional methods such as constant false alarm rate (CFAR) detection and the generalized likelihood ratio test (GLRT) typically construct a detection statistic from a statistical model of the background and compare it with a threshold to make a detection decision~\cite{11267251,9658499,9460779}. The detection performance of these methods depends on how accurately the model characterizes the distribution of the observed background echoes~\cite{10742946}. In complex environments, clutter heterogeneity and nonstationarity can cause the distribution of background echoes to deviate substantially from the assumed distribution, severely degrading detection performance~\cite{9145700}.

Intelligent detectors based on deep learning use neural networks to learn discriminative features of targets and background from labeled echoes and use these features to construct detection statistics~\cite{11676411,11124398,10588477}. Because neural networks can model complex relationships among features nonlinearly and data-driven learning relies less on prespecified background distributions, these detectors can achieve better detection performance than conventional methods in complex clutter environments~\cite{10415070,11278037,10292941}.

However, training intelligent detectors requires accurate, fine-grained labels for individual range cells~\cite{9807335,11204961,10967504}. Obtaining these labels requires precise mapping of ground-truth target positions to discrete range cells, a process that involves coordinate transformations, correction for system delay and range bias, and temporal alignment of target positions with echo acquisition, making annotation costly~\cite{gisinger2021geolocation}. By contrast, segment-level target presence labels are easier to obtain because they indicate only whether a fixed-length echo segment contains a target~\cite{fatseas2022weakly,10945987}. We therefore investigate how to train intelligent detectors using only these segment-level labels, which provide incomplete supervision at the range-cell level.

For an echo segment known to contain a target, the segment-level label does not identify the range cell occupied by the target and therefore cannot be used directly to compute a supervised loss over all range cells in that segment. To reduce reliance on target location annotations, some methods adopt an anomaly detection approach. The range cell being evaluated is referred to as the cell under test (CUT). These methods train a background model using only target-free echoes and determine whether a target is present in the CUT based on how far its echo deviates from the learned background~\cite{rouzoumka2025out,8374078,10636969,RU2026110235}. For example, Bai et al.~\cite{bai2026radar} proposed a detection-oriented complex-valued variational autoencoder (CVVAE). During training, the model learns to reconstruct target-free complex echoes from the same range cell over consecutive pulses. It uses a complex multivariate $t$ likelihood to quantify the fit to the observed echoes and constrains reconstruction errors in the time-domain magnitude and the peaks of the Doppler spectrum, thereby learning the temporal and spectral characteristics of clutter. During detection, the model reconstructs the echo in the CUT and combines the reconstruction errors in the complex echo, its time-domain magnitude, and its Doppler magnitude spectrum to form a detection statistic that measures how far the echo deviates from the learned clutter characteristics. The method achieved good detection performance in experiments with simulated targets superimposed on measured sea clutter. However, because the model is trained only on target-free echoes, it does not use target-containing segments or the segment-level target presence information encoded in their incomplete labels. Leaving this supervision unused limits further gains in detection performance. In addition, the method relies on a slow-time sequence of echoes from multiple pulses, and its detection performance degrades substantially when only a single-pulse echo is available.

\begin{figure}[t]
	\centering
	\includegraphics[width=\columnwidth]{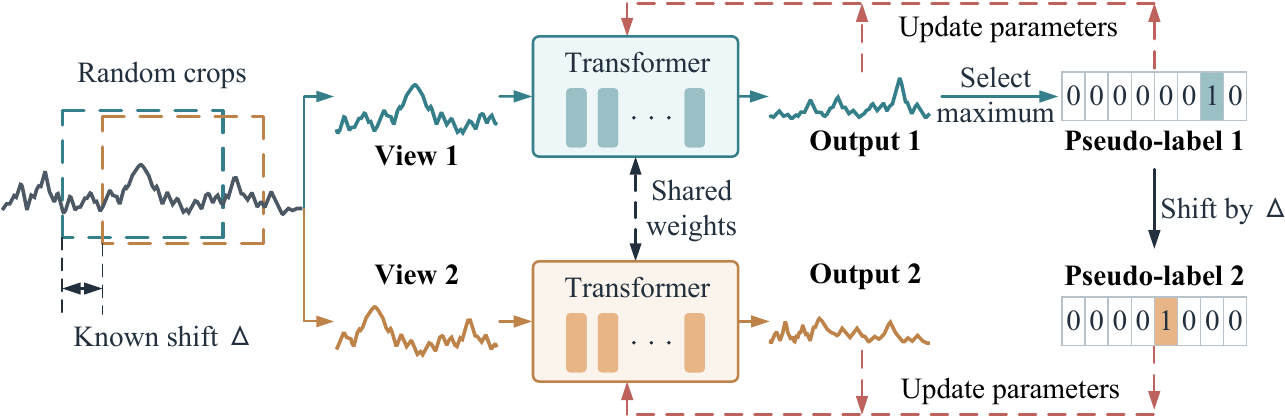}
	\caption{Training procedure for RID-CGL~\cite{11730983}. The detector computes range-cell responses for two random crops of the same echo segment containing a target. The location of the peak response in the first view determines its pseudo-labels, which are mapped to the second view using the known relative displacement $\Delta$.}
	\label{fig_igarss_method}
	\vspace{-0.2cm}
\end{figure}

To make full use of both target-free and target-containing echoes and the supervision provided by their incomplete labels, we previously proposed RID-CGL~\cite{11730983} for training a single-pulse radar detector. As shown in Fig.~\ref{fig_igarss_method}, RID-CGL combines a Siamese Transformer with a translation equivariance constraint along the range dimension. RID-CGL first takes two random crops of the same echo segment to obtain two views with a known relative displacement. The two branches of the Siamese Transformer share parameters and compute a detection response for each range cell in their respective views. RID-CGL then constructs range-cell pseudo-labels from the segment-level target presence label. For a target-free echo segment, it labels all range cells in both views as background. For an echo segment containing a target, it labels the range cell with the highest detection response in the first view as the target and all remaining range cells in that view as background. It then maps the pseudo-labels from the first view to the second using the known displacement between the views. Finally, the pseudo-labels in each view supervise the corresponding detection responses, and both views contribute to a joint update of the detector parameters. This mapping assigns the same label to range cells that correspond to the same physical location in the two views, imposing a translation equivariance constraint along the range dimension during training.

However, RID-CGL selects the target location from the detector's own responses and then uses the resulting pseudo-labels to update the detector. When the signal-to-clutter ratio (SCR) is low or background fluctuations are strong, a background range cell may produce the highest response in an echo segment containing a target. In this case, the background range cell is mislabeled as the target, and the true target range cell is labeled as background. The displacement mapping used to impose the translation equivariance constraint therefore carries this labeling error from the first view to the second. Fitting both sets of incorrect pseudo-labels further increases the detector's response to the mislabeled background range cell and suppresses its response to the true target. The mislabeled location is then more likely to be selected as the target in subsequent iterations, hindering the learning of features that distinguish targets from background and degrading detection performance.

Regarding the use of supervisory information, RID-CGL employs segment-level target presence labels solely to construct range-cell pseudo-labels, which in turn supervise the detection response in each range cell. Even when a pseudo-label places the target in the wrong range cell, the segment-level label still reliably indicates that the echo segment contains a target. The detector nevertheless updates its parameters using only the incorrect location annotation without directly using the correct segment-level label to learn whether the segment contains a target. Regarding training data composition, each echo segment in the original training set contains at most one target, leaving the detector without training examples in which multiple targets occur in the same segment. This makes the detector prone to a single-target bias that limits its ability to detect multiple targets~\cite{zhang2021mosaicos,ding2026grex}. For feature extraction, RID-CGL relies solely on raw echoes, so the detector must learn to distinguish targets from background without direct guidance from background statistics~\cite{oh2021background,10167839,10423036}. In addition, its single convolutional layer extracts local features over a fixed-width range neighborhood, making it difficult to capture both fine details of target echoes and background variations across a wider span of range cells~\cite{rs16183387,ZHANG2021190,10207899}.

To address these limitations, we refine the RID-CGL training method to improve target detection performance. The main contributions are as follows.
\begin{enumerate}
	\item \textit{Background-guided pseudo-label generation}: 
	We first fit a statistical background model to all target-free echo segments in the training set to learn patterns of background fluctuations in individual range cells and correlations between echoes from different range cells. For each echo segment containing a target, we then use the background model to estimate the background component in each range cell. Next, an anomaly score is computed for each range cell to measure how far its observed echo deviates from the estimated background component. Finally, we use the distribution of anomaly scores across the range cells in the segment to estimate the target location and generate the corresponding refined range-cell pseudo-labels. The background model generates these pseudo-labels independently of the detector's current predictions. This prevents the detector's own localization errors from feeding back through the labels and being repeatedly reinforced during subsequent training, helping improve detection performance.
	
	\item \textit{Physically consistent echo augmentation}:
	We linearly superimpose two complex-valued echo segments and generate refined range-cell pseudo-labels and a segment-level target presence label for the augmented echo. Combining two single-target segments yields multitarget samples that reduce the bias introduced by training on segments with at most one target and improve multitarget detection performance. Superimposing two target-free segments creates a more complex clutter background, while adding a target-free segment to a single-target segment makes the target harder to distinguish from clutter. These two pairings further increase the complexity and diversity of the training samples.
	
	\item \textit{Background-aware detector}:
	The detector receives each echo segment along with the whitened residuals and anomaly scores that the background model computes from it. This combination allows the detector to jointly use information from the echo itself and background statistics to distinguish targets from background. Convolutional kernels of different widths then extract multiscale local features from these inputs in parallel, capturing both fine details of the target response across adjacent range cells and clutter variations over wider range neighborhoods. Finally, global attention aggregates local features from across the segment for each range cell, allowing the detector to base its decision in that range cell on both local details of the target echo and background variations throughout the segment. The detector also includes a dedicated classification token that aggregates features from all range cells to capture global information about target presence in the segment.
	
	\item \textit{Multilevel training loss}: 
	We use refined range-cell pseudo-labels to construct a location supervision loss and segment-level target presence labels to construct an auxiliary classification loss and a segment-level presence loss. Jointly optimizing these losses enables training without precise range-cell annotations and guides the detector to learn features that distinguish target echoes from background echoes. Even when the location pseudo-labels in an echo segment containing a target are inaccurate, the segment-level label still provides reliable supervision for target presence, reducing the impact of these errors on training.
\end{enumerate}
The remainder of this paper is organized as follows. 
Section~\ref{sec_problem_formulation_and_related_work} presents the signal model and formulates the training problem under incomplete labeling. Section~\ref{sec_method} describes the proposed method. Section~\ref{sec_experiments} presents the experimental setup and analyzes the results. Section~\ref{sec_conclusion} concludes the paper.

\section{Signal Model and Problem Formulation}\label{sec_problem_formulation_and_related_work}

\subsection{Signal Model}\label{subsec_Signal_Model}
Consider a monostatic radar that transmits a linearly frequency-modulated (LFM) pulse. The transmitted signal is given by
\begin{equation}
	s(t) = {\rm{rect}}\left( {\frac{t}{T_{\text{r}}}} \right)
	\exp \left( {j2\pi \left( {{f_{\text{c}}}t + \frac{1}{2}k{t^2}} \right)} \right),
\end{equation}
where $f_{\text{c}}$ is the carrier frequency, $k$ is the chirp rate of the LFM signal, $T_{\text{r}}$ is the pulse width, and ${\rm{rect}}(\cdot)$ is the rectangular window function:

\begin{equation}
	{\rm{rect}}(u)=
	\begin{cases}
		1, & |u| \leq \frac{1}{2},\\
		0, & \text{otherwise}.
	\end{cases}
\end{equation}
For a point target at range $R$ from the radar, the reflected echo is
\begin{equation}
	r_{\!\text{tar}}(t)
	\!=\!
	\alpha {\rm{rect}}\!\left( \!{\frac{t\!-\!\tau}{T_{\text{r}}}}\! \right)\!
	\exp\!\left(
	\!j2\pi\!\!\left[
	f_{\text{c}}(t\!-\!\tau)
	\!+\!\frac{1}{2}k(t\!-\!\tau)^2
	\right]
	\right)\!,
\end{equation}
where $\tau=2R/c$ is the round-trip propagation delay relative to the transmitted signal, $c$ is the speed of light, and $\alpha$ is a complex coefficient that accounts for target reflectivity and propagation loss.

In addition to the target echo, the received signal contains environmental clutter from nontarget scatterers and receiver noise. The received echo is therefore
\begin{equation}
	r(t) = r_{\text{tar}}(t) + c(t) + n(t), 
\end{equation}
where $c(t)$ denotes environmental clutter and $n(t)$ denotes additive white Gaussian noise. Downconversion, matched filtering, and discrete sampling yield the baseband echo vector $\bm{r}\in\mathbb{C}^{M\times 1}$, where $M$ is the number of range cells~\cite{9540215}. The index of the target range cell is $\lceil\tau f_{\text{s}}\rceil$, where $f_{\text{s}}$ is the sampling rate and $\lceil\cdot\rceil$ denotes the ceiling operation.

Radar target detection determines whether a target echo is present in each range cell. Thus, for any CUT, the detection problem can be expressed as the following binary hypothesis test
\begin{equation}
	\begin{cases}
		H_0: r_m = c_m+n_m,\\
		H_1: r_m = r_{\text{tar},m}+c_m+n_m,
	\end{cases}
	m=1,\ldots,M,
\end{equation}
where $r_m$ denotes the observation in the $m$-th range cell of the echo vector $\bm{r}$. The hypotheses $H_1$ and $H_0$ correspond to target presence and absence, respectively. The CUT is called a target range cell in the former case and a nontarget range cell in the latter.

\subsection{Problem Formulation}\label{subsec_Problem_Formulation}
Conventional detectors typically construct a detection statistic from a prespecified statistical model and compare it with a given threshold $\gamma$ according to the decision rule
\begin{equation}
	\mathcal{T}(r_m)
	\mathop{\gtrless}_{H_0}^{H_1}
	\gamma,
	\label{eq_target_detect}
\end{equation}
where $\mathcal{T}(r_m)$ is the detection statistic for the $m$-th CUT. The detector uses this statistic to distinguish targets from background, and detection performance depends directly on how well the statistic captures the differences between them. In practice, however, clutter often exhibits complex statistical distributions, making it difficult for a statistic derived from a prespecified model to fully capture the differences between target echoes and background interference.

Intelligent detectors use neural networks to automatically extract features from training echoes and learn a nonlinear mapping from echoes to detection statistics. The resulting statistics can better distinguish target range cells from nontarget range cells. Let $f_{\boldsymbol{\theta}}$ denote a detector with parameters $\boldsymbol{\theta}$. For an echo $\bm{r}$, the detector applies the mapping learned during training to produce a detection statistic for each range cell, giving the output vector
\begin{equation}
	f_{\boldsymbol{\theta}}(\bm{r})
	=
	\left[
	f_{\boldsymbol{\theta}}(\bm{r})_1,
	\ldots,
	f_{\boldsymbol{\theta}}(\bm{r})_M
	\right]^{\mathrm{T}}
	\in\mathbb{R}^{M\times 1}.
\end{equation}
To enable $f_{\boldsymbol{\theta}}$ to learn a detection statistic that distinguishes targets from background, most existing methods train $f_{\boldsymbol{\theta}}$ on a dataset $\mathcal{D}_{\mathrm{train}}^{\mathrm{fine}}$ of echo samples paired with accurate range-cell labels:
\begin{equation}
	\begin{gathered}
		\mathcal{D}_{\mathrm{train}}^{\mathrm{fine}}
		=
		\left\{
		\left(
		\bm{r}^{(i)},
		\hat{\mathbf{y}}^{(i)}
		\right)
		\right\}_{i=1}^{N_{\mathrm{train}}^{\mathrm{fine}}}\!\!,
		\quad
		\hat{y}_m^{(i)}
		=
		\begin{cases}
			1, & r_m \in H_1,\\
			0, & r_m \in H_0.
		\end{cases}
	\end{gathered}
\end{equation}
Here, $\mathbf{r}^{(i)}\in\mathbb{C}^{M\times1}$ denotes the $i$-th echo sample, and $\hat{\mathbf{y}}^{(i)}\in\{0,1\}^{M\times1}$ is its vector of accurate range-cell labels. The detector is then trained by minimizing the binary cross-entropy (BCE) loss:
\begin{gather}
	\boldsymbol{\theta}^{\ast}
	=
	\arg\min_{\boldsymbol{\theta}}
	\mathcal{L}_{\mathrm{bce}}(\boldsymbol{\theta}).
	\label{eq_bce_optimization}
	\\
	\begin{aligned}
		\mathcal{L}_{\mathrm{bce}}(\boldsymbol{\theta})
		&=\!
		\frac{-1}{N_{\mathrm{train}}^{\mathrm{fine}}M}\!
		\!\sum_{i=1}^{N_{\mathrm{train}}^{\mathrm{fine}}}
		\!\sum_{m=1}^{M}\!
		\Bigl[
		\hat{y}_m^{(i)}
		\log \sigma\!\left(
		f_{\boldsymbol{\theta}}\left(\bm{r}^{(i)}\right)_m
		\right)
		\\
		&\quad+
		\left(1-\hat{y}_m^{(i)}\right)
		\log\left(
		1-\sigma\left(
		f_{\boldsymbol{\theta}}\left(\bm{r}^{(i)}\right)_m
		\right)
		\right)
		\Bigr].
	\end{aligned}
	\label{eq_bce_loss}
\end{gather}
Here, $\boldsymbol{\theta}^{\ast}$ denotes the optimal parameters obtained through training. Precise range-cell-level labels provide explicit location supervision. The supervised loss constructed from these labels therefore constrains the prediction in each range cell, guiding the detector to learn features that distinguish target echoes from background interference and to construct a corresponding detection statistic from these features.

\begin{figure*}[t]
	\centering
	\includegraphics[width=\textwidth]{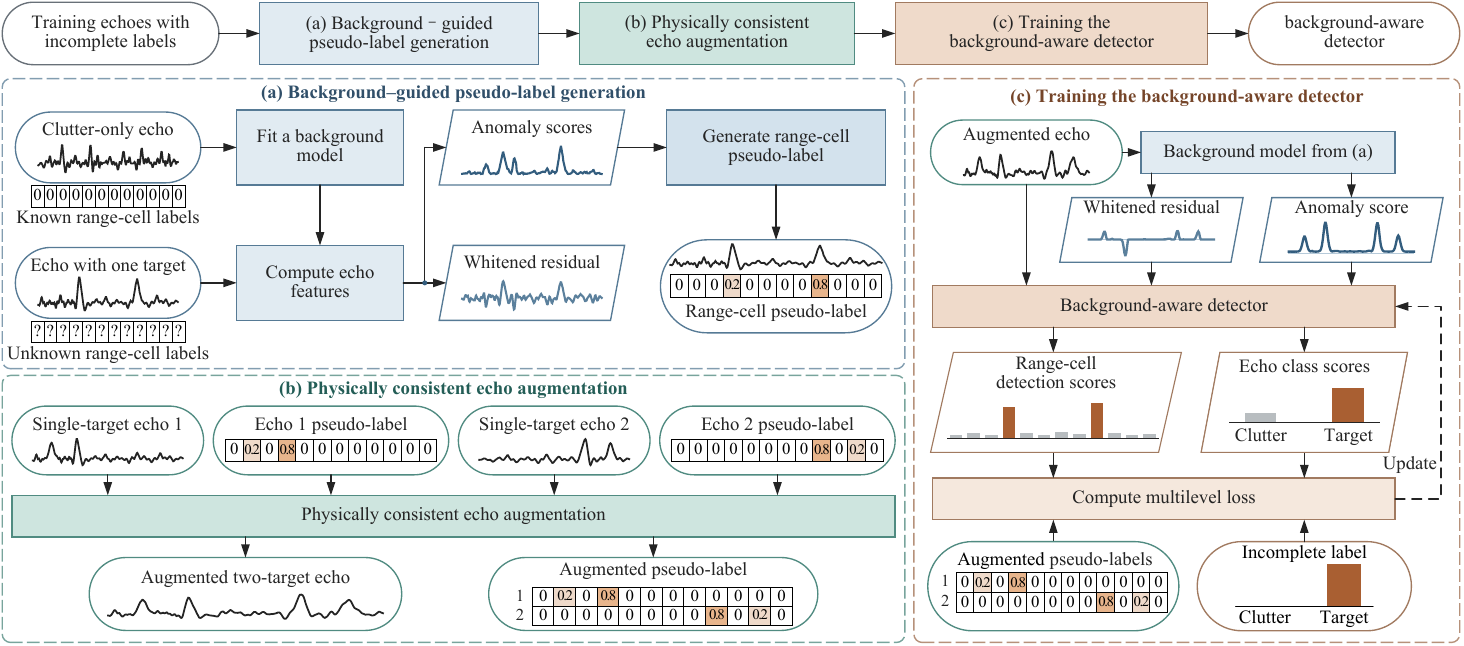}
	\vspace{-0.6cm}
	\caption{Overview of the proposed method for training intelligent radar target detectors using only incomplete labels.}
	\label{fig_method}
\end{figure*}

However, the high cost of obtaining accurate range-cell labels in practice limits the applicability of this training scheme. Determining whether an echo segment contains a target is generally easier than identifying the exact range cell containing the target. To construct such incomplete labels for detector training, RID-CGL~\cite{11730983} partitions the echo $\bm{r}$ along the range dimension into fixed-length segments $\mathbf{x}\in\mathbb{C}^{K\times1}$, each comprising $K$ range cells, and constructs the training set
\begin{equation}
	\begin{aligned}
		{\mathcal{D}}_{\mathrm{train}}
		&=
		\left\{
		\left(
		\mathbf{x}^{(i)},
		y^{(i)}
		\right)
		\right\}_{i=1}^{N_{\mathrm{train}}}, \quad y^{(i)}\in\{0,1\},
	\end{aligned}
\end{equation}
in which each original training segment contains at most one target. The segment-level target presence label is $y^{(i)}=1$ if segment $\mathbf{x}^{(i)}$ contains a target and $y^{(i)}=0$ otherwise. Because the incomplete labels do not specify the precise target location along the range dimension, the training method described above cannot be applied directly to this dataset. We therefore investigate how to make full use of the limited supervision provided by incomplete labels to train a detector $f_{\boldsymbol{\theta}}$ with high detection performance.

\section{Methodology}\label{sec_method} 
\subsection{Overview}
Fig.~\ref{fig_method} shows the proposed training procedure. We first generate refined range-cell pseudo-labels for all training echo segments. Each range cell in a clutter-only segment receives a pseudo-label of zero. We then fit a background model to all clutter-only echo segments in the training set to capture background fluctuations within individual range cells and correlations between echoes from different range cells. For each segment containing a target, we use the fitted model to estimate the background component in each range cell from echoes in other range cells selected as references. Subtracting each estimate from the corresponding observed echo gives the estimation residual for that range cell. After whitening, each residual yields an anomaly score that measures how far the observed echo deviates from the estimated background. The refined pseudo-label marks the two highest-scoring range cells as candidate locations for the same target, reducing the risk of omitting its true location.

Next, we expand the training set through physically consistent echo augmentation. By linearly superimposing the complex echoes from two single-target segments, we can generate training examples with multiple targets in the same segment to improve the detector's performance in multitarget scenarios. In addition, to increase the complexity and diversity of the training examples, we include clutter-only echo segments in the superposition step. Superimposing two clutter-only segments creates a more complex clutter background, while adding a clutter-only segment to a single-target segment makes the target harder to distinguish from the background. For each augmented example, the refined pseudo-labels of the two original segments are stacked in separate rows to preserve each target's candidate locations and weights.

Finally, we train the background-aware detector on the augmented examples. For each augmented echo, the fitted background model provides whitened residuals and anomaly scores that characterize deviations from the estimated background. The detector takes these cues and the augmented echo as inputs, combining background information with echo morphology to learn features that distinguish target echoes from clutter. From these features, it produces a detection score for each range cell and an echo class score indicating whether the segment contains a target. The multilevel loss combines target presence information from the segment-level labels with location cues from the refined range-cell pseudo-labels to guide the detector in identifying range cells that contain a target.

\subsection{Background-Guided Pseudo-Label Generation}

\subsubsection{Background Modeling}
For echo segment $\mathbf{x}^{(n)}$, we arrange its in-phase and quadrature components into the real-valued matrix
\begin{equation}
	\mathbf{X}^{(n)}=
	[\operatorname{Re}(\mathbf{x}^{(n)}),
	\operatorname{Im}(\mathbf{x}^{(n)})]
	\in\mathbb{R}^{K\times2}.
\end{equation}
We refer to echo segments with and without targets as target segments and background segments, respectively. The corresponding sets of training sample indices and echo matrices are
\begin{equation}
	\begin{gathered}
	\mathcal{N}_-=\{n:y^{(n)}=0\},\quad
	\mathcal{D}_{\mathrm{train}}^-=
	\{\mathbf{X}^{(n)}:n\in\mathcal{N}_-\},\\
	\mathcal{N}_+=\{n:y^{(n)}=1\},\quad
	\mathcal{D}_{\mathrm{train}}^+=
	\{\mathbf{X}^{(n)}:n\in\mathcal{N}_+\},
	\end{gathered}
\end{equation}
with $|\mathcal{N}_+|$ target segments and $|\mathcal{N}_-|$ background segments. We fit the background model using all segments in $\mathcal{D}_{\mathrm{train}}^-$ to learn the statistical properties of clutter and noise.

To reduce the impact of differences in overall power across echo segments on background modeling while preserving the phase relationships and relative amplitude structure within each segment, we normalize each echo segment based on its own mean power as follows:
\begin{equation}
	(\nu^{(n)})^2=
	\frac{1}{K}\sum_{i=1}^{K}|x_i^{(n)}|^2,\quad
	\bar{\mathbf{X}}^{(n)}=
	\mathbf{X}^{(n)}/\nu^{(n)}.
	\label{eq_norm}
\end{equation}
Next, to facilitate joint modeling of the statistical relationships among the in-phase and quadrature components across range cells, we stack the columns of the normalized matrix to form a vector:
\begin{equation}
	\mathbf{v}^{(n)}=\operatorname{vec}(\bar{\mathbf{X}}^{(n)})
	\in\mathbb{R}^{2K\times1}.
\end{equation}
Entries $i$ and $K+i$ of $\mathbf{v}^{(n)}$ correspond to the in-phase and quadrature components of range cell $i$, respectively. We model the normalized background echoes using a multivariate Gaussian distribution $\mathcal{N}(\boldsymbol{\mu},\boldsymbol{\Sigma})$, whose parameters are estimated as
\begin{equation}
	\begin{gathered}
	\boldsymbol{\mu}=
	\frac{1}{|\mathcal{N}_-|}\sum_{n\in\mathcal{N}_-}\mathbf{v}^{(n)},\\
	\boldsymbol{\Sigma}=
	\frac{1}{|\mathcal{N}_-|}\sum_{n\in\mathcal{N}_-}
	(\mathbf{v}^{(n)}-\boldsymbol{\mu})
	(\mathbf{v}^{(n)}-\boldsymbol{\mu})^\top.
	\end{gathered}
	\label{eq_background_model}
\end{equation}
The mean vector $\boldsymbol{\mu}$ represents the average background level of each component, while the sample covariance matrix $\boldsymbol{\Sigma}$ characterizes the component variances and the covariances between components.

\subsubsection{Anomaly Score Calculation}	
We first use the background model to estimate the background component in each range cell. We then compute the anomaly score for each range cell from the difference between its observed echo and the corresponding background estimate. Target echo energy extends over the target center and neighboring range cells. Using these neighboring range cells as reference cells may introduce target returns into the background estimate for the CUT, reducing estimation accuracy~\cite{4200717}. To mitigate this effect, we place a guard region around the CUT, as in CFAR detectors, and estimate its background component using only echoes from the reference cells.

Given a guard radius $h_{\mathrm{p}}$ and a reference radius $h_{\mathrm{r}}$, the set of reference cells for the $i$-th CUT is defined as
\begin{equation}
	\mathcal{R}_i=\{j\in\{1,\ldots,K\}:
	h_{\mathrm{p}}<|j-i|\le h_{\mathrm{r}}\},
\end{equation}
and the index sets of the components in $\mathbf{v}^{(n)}$ corresponding to the CUT and its reference cells are, respectively,
\begin{equation}
	\mathcal{T}_i=\{i,K+i\},\quad
	\mathcal{V}_i=\bigcup_{j\in\mathcal{R}_i}\{j,K+j\}.
\end{equation}
We use $\mathcal{V}_i$ to extract the reference-cell observation vector $\mathbf{v}^{(n)}_{\mathcal{V}_i}$ from $\mathbf{v}^{(n)}$. Using $\mathcal{T}_i$ and $\mathcal{V}_i$, we extract the mean vectors $\boldsymbol{\mu}_{\mathcal{T}_i}$ and $\boldsymbol{\mu}_{\mathcal{V}_i}$ for the CUT and its reference cells, respectively, from the background mean $\boldsymbol{\mu}$. Under the joint Gaussian background model, the in-phase and quadrature background components of the $i$-th CUT can be estimated from the reference-cell observation vector using the conditional mean:
\begin{equation}
	\boldsymbol{\mu}_i^{(0)}(\bar{\mathbf{X}}^{(n)})=
	\boldsymbol{\mu}_{\mathcal{T}_i}
	+\boldsymbol{\Sigma}_{\mathcal{T}_i,\mathcal{V}_i}
	\boldsymbol{\Sigma}_{\mathcal{V}_i,\mathcal{V}_i}^{-1}
	(\mathbf{v}^{(n)}_{\mathcal{V}_i}
	-\boldsymbol{\mu}_{\mathcal{V}_i}).
	\label{eq_conditional_background_prediction}
\end{equation}
This expression shows that background estimation requires inverting the reference-cell sample covariance submatrix $\boldsymbol{\Sigma}_{\mathcal{V}_i,\mathcal{V}_i}$. However, a limited number of training samples or strong correlations among the reference echo components may make this matrix nearly singular. Inverting it can then amplify estimation errors and degrade the background estimate~\cite{10214148}. To stabilize the inversion, we apply shrinkage to the sample covariance in~\eqref{eq_background_model} and add a diagonal regularization term to ensure that the resulting matrix is positive definite:
\begin{equation}
	\tilde{\boldsymbol{\Sigma}}=
	(1-\eta)\boldsymbol{\Sigma}
	+\eta\zeta\mathbf{I}_{2K}+\epsilon^2\mathbf{I}_{2K},\quad
	\zeta=\frac{\operatorname{tr}(\boldsymbol{\Sigma})}{2K}.
	\label{eq_shrinkage_estimation}
\end{equation}
Here, $\eta\in[0,1]$ is the shrinkage coefficient, $\epsilon>0$ is a numerical stabilization parameter, and $\mathbf{I}_{2K}$ is the identity matrix. Covariance shrinkage draws the component variances toward their mean and attenuates the off-diagonal covariance terms, thereby mitigating inversion instability caused by limited training samples and strong correlations among components~\cite{boeck2025gohberg}. Substituting $\tilde{\boldsymbol{\Sigma}}$ for $\boldsymbol{\Sigma}$ in~\eqref{eq_conditional_background_prediction} then yields the following estimate of the background components:
\begin{equation}
	\begin{gathered}
	\mathbf{B}_i=
	\tilde{\boldsymbol{\Sigma}}_{\mathcal{T}_i,\mathcal{V}_i}
	\tilde{\boldsymbol{\Sigma}}_{\mathcal{V}_i,\mathcal{V}_i}^{-1},\\
	\boldsymbol{\mu}_i(\bar{\mathbf{X}}^{(n)})=
	\boldsymbol{\mu}_{\mathcal{T}_i}
	+\mathbf{B}_i(\mathbf{v}^{(n)}_{\mathcal{V}_i}
	-\boldsymbol{\mu}_{\mathcal{V}_i}).
	\end{gathered}
\end{equation}

For the normalized echo segment $\bar{\mathbf{X}}^{(n)}$, the estimation residual in the $i$-th range cell is
\begin{equation}
	\mathbf{r}_{\mathrm{res},i}(\bar{\mathbf{X}}^{(n)})=
	(\bar{\mathbf{X}}^{(n)}_{i,:})^\top
	-\boldsymbol{\mu}_i(\bar{\mathbf{X}}^{(n)}).
\end{equation}
Under background conditions, the in-phase and quadrature estimation residuals may have different variances and be correlated. Consequently, residuals of the same magnitude need not indicate the same degree of anomaly. To account for the variances and correlation of the background estimation errors when computing the anomaly score, we first compute the regularized conditional covariance matrix:
\begin{equation}
	\mathbf{S}_i=
	\tilde{\boldsymbol{\Sigma}}_{\mathcal{T}_i,\mathcal{T}_i}
	-\mathbf{B}_i\tilde{\boldsymbol{\Sigma}}_{\mathcal{V}_i,\mathcal{T}_i}
	+\epsilon^2\mathbf{I}_2.
\end{equation}
We then perform a Cholesky decomposition of $\mathbf{S}_i$ and use the inverse of the resulting lower-triangular factor to whiten the estimation residual:
\begin{equation}
	\mathbf{S}_i=\mathbf{A}_i\mathbf{A}_i^\top,\quad \mathbf{e}_i(\bar{\mathbf{X}}^{(n)})=
	\mathbf{A}_i^{-1}
	\mathbf{r}_{\mathrm{res},i}(\bar{\mathbf{X}}^{(n)}),
\end{equation}
where $\mathbf e_i$ is the whitened residual. Whitening standardizes the in-phase and quadrature estimation residuals relative to the variability of background estimation errors and decorrelates the two components. The two whitened residual components can thus be compared on a common scale, and the sum of their squares gives the anomaly score for the $i$-th range cell,
\begin{equation}
	s_i(\bar{\mathbf{X}}^{(n)})=
	\mathbf{e}_i(\bar{\mathbf{X}}^{(n)})^\top
	\mathbf{e}_i(\bar{\mathbf{X}}^{(n)}).
\end{equation}
A higher anomaly score indicates a greater deviation of the observed echo in that range cell from the estimated background.

\subsubsection{Pseudo-Label Generation} 
Let $\mathbf{q}^{(n)}=[q_1^{(n)},\ldots,q_K^{(n)}]^\top$ denote the vector of refined range-cell pseudo-labels for $\bar{\mathbf{X}}^{(n)}$. For each background segment, we set $\mathbf{q}^{(n)}$ to the zero vector. For each target segment, since it contains only one target, we can estimate the target's range cell as the one with the highest anomaly score and construct the corresponding pseudo-label as follows:
\begin{equation}
	k_{\max}^{(n)}=
	\argmax_{1\le i\le K}s_i(\bar{\mathbf{X}}^{(n)}),\quad
	q_i^{(n)}=
	\begin{cases}
		1,&i=k_{\max}^{(n)},\\
		0,&i\ne k_{\max}^{(n)}.
	\end{cases}
\end{equation}

However, background fluctuations may cause a background range cell to have a higher anomaly score than the target range cell, potentially leading to the latter being mislabeled as background. To reduce this risk, we rank the range cells in descending order of their anomaly scores and select the top $K_{\mathrm c}$ range cells as follows:
\begin{equation}
	(k_1^{(n)},\ldots,k_{K_{\mathrm c}}^{(n)})
	=\operatorname{Top}_{K_{\mathrm c},\,1\le i\le K}
	s_i(\bar{\mathbf{X}}^{(n)}).
\end{equation}
These range cells are candidate locations for the target. We then assign location weights according to the anomaly-score ranking, with a higher weight indicating that the target is more likely to occupy the corresponding range cell:
\begin{equation}
	\begin{gathered}
	q_{k_1^{(n)}}^{(n)}\ge\cdots\ge
	q_{k_{K_{\mathrm c}}^{(n)}}^{(n)}>0,\quad
	\sum_{r=1}^{K_{\mathrm c}}q_{k_r^{(n)}}^{(n)}=1,\\
	q_i^{(n)}=0,\quad
	i\notin\{k_r^{(n)}:r=1,\ldots,K_{\mathrm c}\}.
	\end{gathered}
\end{equation}
Thus, even if the range cell with the highest anomaly score does not contain the target, another candidate may still contain it, reducing the risk of mislabeling the target's range cell as background.

\subsection{Physically Consistent Echo Augmentation}

Let $p\in(0,1)$ denote the proportion of target segments in the training set. We call the echo segment drawn according to the training class proportions the base segment and the echo segment selected for superposition the donor segment. To increase the number of multitarget training examples, we always apply superposition when the base is a target segment. In this case, we draw the donor from the target or background segment pool with probabilities $(3-p)/4$ and $(1+p)/4$, respectively. When the base is a background segment, we apply superposition with probability 0.5 and draw the donor uniformly from the entire training set. Otherwise, the base segment is left unaugmented. This strategy increases the proportion of target-containing training examples after augmentation and ensures that single-target and multitarget cases each account for half of these examples, thereby balancing the two training scenarios. Appendix~\ref{proof_1} gives the proof.

Given a base segment $\mathbf{X}^{(a)}$ and a donor segment $\mathbf{X}^{(b)}$, let $\delta^{(a)}\in\{0,1\}$ indicate whether superposition is applied. The augmented echo segment is
\begin{equation}
	\tilde{\mathbf{X}}^{(a)}=
	\begin{cases}
		\mathbf{X}^{(a)},&\delta^{(a)}=0,\\[1mm]
		(\mathbf{X}^{(a)}+\mathbf{X}^{(b)})/\sqrt{2},
		&\delta^{(a)}=1.
	\end{cases}
\end{equation}
Because superposition preserves target presence in the constituent segments, we take the logical OR of their segment-level labels to obtain the augmented segment-level label. We concatenate the refined pseudo-labels of the base and donor segments along the row dimension to retain the candidate range cells and their weights for each target:
\begin{equation}
	\tilde y^{(a)}
	\!=\!
	\left\{
	\begin{array}{@{}l@{\,}l@{}}
		y^{(a)}, & \delta^{(a)}=0,\\
		y^{(a)}\lor y^{(b)}, & \delta^{(a)}=1,
	\end{array}
	\right.
	\quad
	\mathbf{Q}_{\mathrm c}
	\!=\!
	\begin{bmatrix}
		(\mathbf{q}^{(a)})^\top\\
		\delta^{(a)}(\mathbf{q}^{(b)})^\top
	\end{bmatrix}.
\end{equation}

\begin{figure*}[t]
	\centering
	\includegraphics[width=\textwidth]{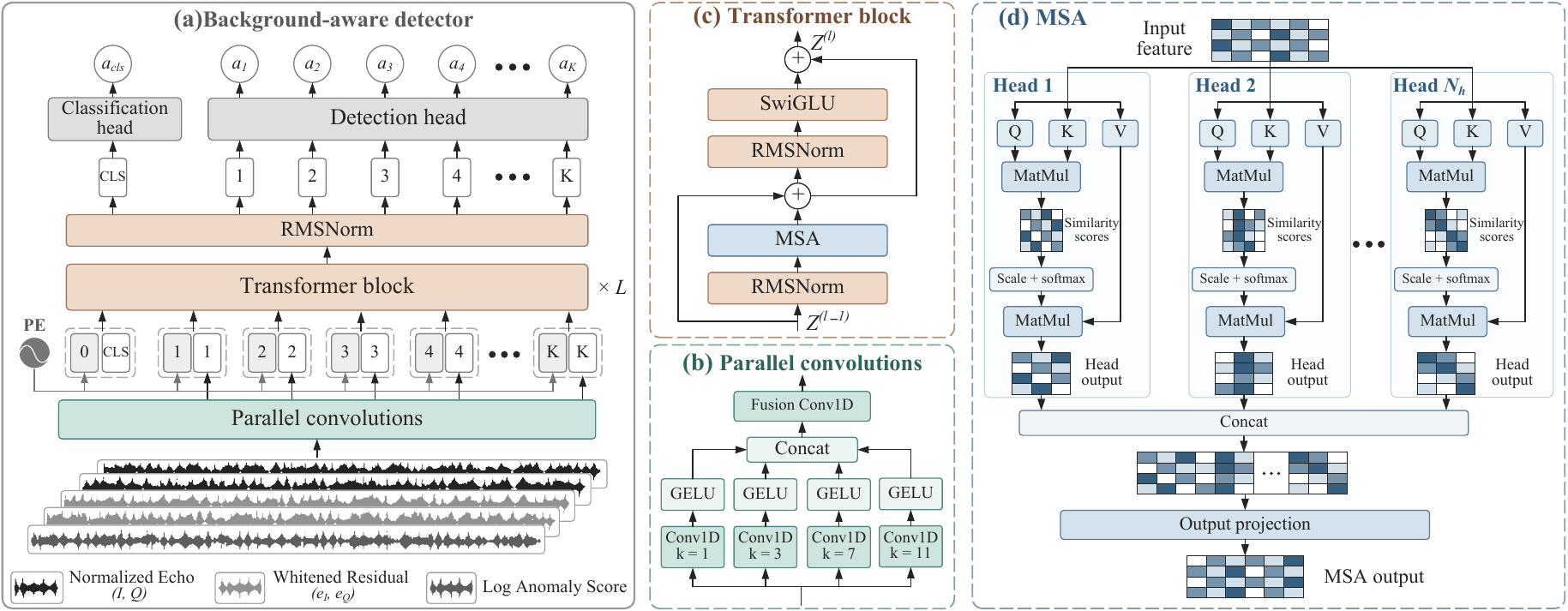}
	\vspace{-0.6cm}
	\caption{Architecture and main components of the background-aware detector.}
	\label{fig_model}
\end{figure*}

\subsection{Background-Aware Detector}
As shown in Fig.~\ref{fig_model}, the background-aware detector takes the normalized echo together with the whitened residuals and log anomaly scores computed by the background model as inputs to make full use of background statistics in guiding target detection. Parallel convolutions then extract multiscale local features for each range cell from these inputs. Multihead self-attention (MSA) subsequently integrates background information from the entire echo segment into the local features of each range cell, enabling the detector to distinguish target echoes from background fluctuations.

\subsubsection{Input Construction}
The augmented echo segment $\tilde{\mathbf{X}}$ is power-normalized according to~\eqref{eq_norm} to obtain $\bar{\mathbf{X}}$. The background model then computes the whitened residuals $\mathbf{e}(\bar{\mathbf{X}})$ and anomaly scores $s(\bar{\mathbf{X}})$ from the normalized echo. Concatenating the normalized echo, whitened residuals, and anomaly scores yields a five-channel feature matrix $\mathbf{F}\in\mathbb{R}^{K\times5}$, where the $i$-th row $\mathbf{F}_{i,:}$ represents the input features for the $i$-th range cell:
\begin{equation}
	\mathbf{F}_{i,:}=
	[\bar x_{i,I},
	\bar x_{i,Q},
	e_{i,I},e_{i,Q},\log(1+s_i)],
\end{equation}
where the subscripts $I$ and $Q$ denote the in-phase and quadrature components, respectively. The normalized echo preserves phase relationships and relative amplitude structure. The whitened residuals provide component-wise information on deviations from the conditional background estimate, while the log-transformed anomaly score quantifies the deviation magnitude and compresses its dynamic range. Even when a target echo does not stand out in amplitude, the detector can use this background information to distinguish weak targets from clutter fluctuations, thereby improving the detection probability of weak targets.

\subsubsection{Parallel Convolutions}
To encode both the local details of each range cell and the echo structure in its neighborhood, we apply four parallel one-dimensional convolutional branches to the input echo features $\mathbf{F}$:
\begin{equation}
	\mathbf{H}^{(\kappa)}=
	\operatorname{GELU}\!\left(\operatorname{BN}_{\kappa}
	(\operatorname{Conv}_{\kappa}(\mathbf{F}))\right),
	\ \kappa\in\{1,3,7,11\}.
\end{equation}
Here, $\kappa$ denotes the convolutional kernel width, $d_{\mathrm b}$ is the number of output channels in each branch, and $\operatorname{BN}_{\kappa}$ denotes batch normalization in the corresponding branch. The kernel of width $1$ captures echo details within a single range cell, while the other kernels extract echo features from neighborhoods at different scales.

We then concatenate the branch outputs along the channel dimension and fuse them using a pointwise convolution with a kernel width of 1 to obtain the multiscale feature matrix $\mathbf{H}$:
\begin{equation}
	\mathbf{h}_i\!=\mathbf{H}_{i,:}^\top, \quad \mathbf{H}=\operatorname{Conv}^{\mathrm f}_{1}\!\left(
	\operatorname*{Concat}_{\kappa\in\{1,3,7,11\}}
	\mathbf{H}^{(\kappa)}\right)\in\mathbb{R}^{K\times d}.
\end{equation}
Here, $\mathbf{h}_i$ denotes the multiscale feature of the $i$-th range cell.

\subsubsection{Global Attention}
We prepend a global classification token (CLS) $\mathbf{h}_{\mathrm{cls}}\in\mathbb{R}^{d}$ to the multiscale feature sequence. Through global attention, this token aggregates features from all range cells, enabling the classification head to determine whether the echo segment contains a target. To incorporate positional relationships among range cells into global attention, we add the positional encoding $\mathbf{E}\in\mathbb{R}^{(K+1)\times d}$ to form the global feature sequence:
\begin{equation}
	\mathbf{Z}^{(0)}=
	[\mathbf{h}_{\mathrm{cls}},\mathbf{h}_1,\ldots,\mathbf{h}_K]^\top
	+\mathbf{E}.
\end{equation}
The resulting sequence is then passed through $L$ successive Transformer blocks, allowing the local representation of each range cell to progressively integrate information from the entire echo segment:
\begin{equation}
	\begin{gathered}
	\mathbf{U}=\operatorname{RMSNorm}(\mathbf{Z}^{(\ell-1)}), \ \tilde{\mathbf{Z}}^{(\ell)}=
	\mathbf{Z}^{(\ell-1)}\!+\!
	\operatorname{MSA}(\mathbf{U}),\\
	\mathbf{Z}^{(\ell)}=\tilde{\mathbf{Z}}^{(\ell)}+
	\operatorname{SwiGLU}(\operatorname{RMSNorm}(\tilde{\mathbf{Z}}^{(\ell)})), 
	\end{gathered}
\end{equation}
where $\ell\in\{1,\ldots,L\}$.

MSA comprises $N_{\mathrm h}$ attention heads. Each head linearly projects the input features into queries, keys, and values and aggregates information according to the feature similarity between different range cells. For attention head $j$, the projection and aggregation are given by
\begin{equation}
	\begin{gathered}
	\bigl[\mathbf{Q}^{(j)},\mathbf{K}^{(j)},\mathbf{V}^{(j)}\bigr]
	=\mathbf{U}\bigl[\mathbf{W}_{Q}^{(j)},\mathbf{W}_{K}^{(j)},\mathbf{W}_{V}^{(j)}\bigr],\\
	\mathbf{A}^{(j)}=
	\operatorname{softmax}\!\left(
	\frac{\mathbf{Q}^{(j)}(\mathbf{K}^{(j)})^\top}
	{\sqrt{d_h}}\right)\mathbf{V}^{(j)}.
	\end{gathered}
\end{equation}
Each entry of $\mathbf{Q}^{(j)}(\mathbf{K}^{(j)})^\top$ reflects the feature similarity between the two corresponding range cells. Scaling these scores and applying softmax yields the attention weights. Using these weights, each range cell aggregates the value components in $\mathbf{V}^{(j)}$ corresponding to all range cells. Its local representation thus incorporates background information from the entire echo segment, helping distinguish target echoes from local anomalies caused by background fluctuations. We then concatenate the head outputs along the channel dimension and apply a linear projection to obtain the MSA output:	
\begin{equation}
	\operatorname{MSA}(\mathbf{U})=
	\left(\operatorname*{Concat}_{j=1,\ldots,N_{\mathrm h}}
	\mathbf{A}^{(j)}\right)\mathbf{W}_{O}\in\mathbb{R}^{(K+1)\times d}.
\end{equation}

After the $L$ Transformer blocks, we apply RMSNorm to the resulting global feature sequence to normalize the scale of each feature vector, making the output scores less sensitive to differences in vector magnitude. The detection and classification heads then produce range-cell-level and segment-level outputs, respectively:
\begin{equation}
	\begin{gathered}
	\hat{\mathbf{Z}}=\operatorname{RMSNorm}(\mathbf{Z}^{(L)}),\quad
	\hat{\mathbf{z}}_i=\hat{\mathbf{Z}}_{i,:}^\top,\\
	a_i=\mathbf{w}_{\mathrm{cell}}^\top\hat{\mathbf{z}}_i+b_{\mathrm{cell}},\quad
	a_{\mathrm{cls}}=\mathbf{w}_{\mathrm{cls}}^\top\hat{\mathbf{z}}_0+b_{\mathrm{cls}}.
	\end{gathered}
\end{equation}
Here, $a_i$ denotes the detection statistic for the corresponding range cell. The vector $\hat{\mathbf{z}}_0$ aggregates features from all range cells to form a global representation of the echo segment. The classification head uses this representation to produce the detection statistic $a_{\mathrm{cls}}$, which indicates whether a target is present in the echo segment.

\subsection{Multilevel Loss}

To formulate the loss terms, we first convert the detection statistics into scores using the sigmoid function:
\begin{equation}
	p_i\!=\!\sigma(a_i),\
	p_{\mathrm{cls}}\!=\!\sigma(a_{\mathrm{cls}}),\ \sigma(\xi)\!=\!(1+\exp(-\xi))^{-1},
\end{equation}
where $p_{\mathrm{cls}}$ denotes the segment-level target presence score produced by the classification head, and $p_i$ denotes the detection score for the $i$-th range cell. The vector of range-cell detection scores is denoted by $\mathbf{p}=[p_1,\ldots,p_K]^\top$.

\subsubsection{Classification Loss}
We use the augmented segment-level label $\tilde y$ to supervise the target presence prediction of the classification head. The corresponding BCE loss is
\begin{equation}
	\ell_{\mathrm{cls}}(p_{\mathrm{cls}},\tilde y)=
	-\tilde y\log p_{\mathrm{cls}}
	-(1-\tilde y)\log(1-p_{\mathrm{cls}}).
\end{equation}
Gradients from this loss propagate through the classification head and global attention layers to the weights of the parallel convolutions. The segment-level labels can thus guide the learning of both local and global features, helping the detector distinguish targets from background in each range cell.

\subsubsection{Background Segment Loss}
For a background segment with augmented label $\tilde y=0$, the refined pseudo-label marks every range cell as background. We therefore constrain the detection head outputs across all range cells using the following BCE loss, which guides the detector to learn background features associated with clutter and noise and suppress spurious target responses:
\begin{equation}
	\ell_{\mathrm{neg}}(\mathbf{p})=
	-\frac{1}{K}\sum_{i=1}^{K}\log(1-p_i).
\end{equation}

\subsubsection{Refined Pseudo-Label Loss}
For an augmented target segment, the refined pseudo-labels provide candidate range cells for each target along with their corresponding weights. To exploit these location cues, we aggregate the range-cell detection scores using the corresponding candidate weights and construct a BCE loss:
\begin{equation}
	\begin{gathered}
		m_t=\sum_{i=1}^{K}(\mathbf{Q}_{\mathrm c})_{t,i},
		\quad
		u_t=\sum_{i=1}^{K}(\mathbf{Q}_{\mathrm c})_{t,i}p_i,
		\quad t=1,2,\\
		\ell_{\mathrm{src}}^{\mathrm{bce}}(\mathbf{p},\mathbf{Q}_{\mathrm c})
		\!=\!\frac{-1}{m_1\!+\! m_2}
		\!\sum_{t=1}^{2}\!\mathbf{1}[m_t>0]
		\log\!\left(u_t\!+\!1\!-\!m_t\right),
	\end{gathered}
	\label{eq_bce1}
\end{equation}
where $\mathbf{1}[\cdot]$ denotes the indicator function, $u_t$ denotes the weighted aggregate of range-cell detection scores, and $m_t$ indicates whether row $t$ contains a target.

Constraining target responses only at candidate locations leaves spurious target responses in background range cells outside the candidate set unconstrained. We therefore let $\mathcal{U}=\{i:(\mathbf{Q}_{\mathrm c})_{1,i}+(\mathbf{Q}_{\mathrm c})_{2,i}>0\}$ denote the union of candidate locations and apply the following BCE loss to the range cells labeled as background by the refined pseudo-labels:
\begin{equation}
	\ell_{\mathrm{off}}^{\mathrm{bce}}(\mathbf{p},\mathbf{Q}_{\mathrm c})=
	-\sum_{i\notin\mathcal{U}}\log(1-p_i).
	\label{eq_bce2}
\end{equation}
This loss term constrains the detection scores of range cells outside the candidate set to reduce false alarms caused by misclassifying background range cells as targets.

However, refined pseudo-labels generated by the background model may miss true target locations, incorrectly labeling target range cells as background and background range cells as targets. If training continues with $\ell_{\mathrm{src}}^{\mathrm{bce}}$ and $\ell_{\mathrm{off}}^{\mathrm{bce}}$, these errors can encourage the detector to increase target responses in background range cells incorrectly labeled as targets and suppress responses in target range cells incorrectly labeled as background, thereby interfering with the learning of features that distinguish targets from background~\cite{9467531,9585690}. To mitigate this effect, we expand the negative logarithmic term $-\log \xi$ in $\ell_{\mathrm{src}}^{\mathrm{bce}}$ and $\ell_{\mathrm{off}}^{\mathrm{bce}}$ in a Taylor series about $\xi=1$ and retain the first four terms. This yields the fourth-order Taylor cross-entropy (TCE) loss
\begin{equation}
	\ell_{\mathrm{TCE}}^{(4)}(\xi)=\sum_{r=1}^{4}\frac{(1-\xi)^r}{r},
	\qquad \xi\in[0,1],
	\label{eq_tce}
\end{equation}
which replaces the negative logarithmic term in $\ell_{\mathrm{src}}^{\mathrm{bce}}$ and $\ell_{\mathrm{off}}^{\mathrm{bce}}$. Truncation to four terms keeps $\ell_{\mathrm{TCE}}^{(4)}(\xi)$ finite as $\xi$ approaches zero, thereby limiting the penalty when predictions strongly disagree with pseudo-labels and reducing the adverse effect of erroneous pseudo-labels on the learning of detection features~\cite{feng2021can}.

Combining~\eqref{eq_bce1},~\eqref{eq_bce2}, and~\eqref{eq_tce} gives the refined pseudo-label loss
\begin{equation}
	\begin{gathered}
		\ell_{\mathrm{fine}}(\mathbf{p},\mathbf{Q}_{\mathrm c})
		=\frac{1}{m_1+m_2}
		\sum_{t=1}^{2}\mathbf{1}[m_t>0]\ell_{\mathrm{TCE}}^{(4)}(u_t)\\
		+\lambda_{\mathrm{off}}
		\sum_{i\notin\mathcal{U}}\ell_{\mathrm{TCE}}^{(4)}(1-p_i),
	\end{gathered}
\end{equation}
where $\lambda_{\mathrm{off}}$ weights the suppression term for range cells outside the candidate set. Because nontarget range cells far outnumber target range cells in a target segment, we set $\lambda_{\mathrm{off}}=0.05$ to reduce the risk of the accumulated background loss terms overwhelming the supervision at candidate target locations.

\subsubsection{Presence Loss}
The incomplete label $\tilde y = 1$ of an augmented target segment indicates that the echo segment contains at least one target. To incorporate this information into range-cell-level detection training, we base the presence loss on a smooth maximum of the range-cell detection statistics:
\begin{equation}
	\begin{gathered}
		g_{\tau}(\mathbf{a})=
		\tau\log\left(
		\frac{1}{K}\sum_{i=1}^{K}
		\exp\left(\frac{a_i}{\tau}\right)
		\right),\\
		h(\mathbf{a})=\sigma(g_{\tau}(\mathbf{a})),\quad
		\ell_{\mathrm{pre}}(\mathbf{a})=-\log h(\mathbf{a}),
	\end{gathered}
\end{equation}
where $\mathbf{a}=[a_1,\ldots,a_K]^\top$ is the vector of range-cell detection statistics, and $g_{\tau}(\mathbf{a})$ is the pooled segment-level detection statistic that combines range-cell responses to represent target presence across the segment. The parameter $\tau>0$ controls the smoothness of the aggregation and is set to $0.4$. $h(\mathbf{a})$ denotes the segment-level target presence score. Since $h(\mathbf{a})\leq\max_i p_i$, $h(\mathbf{a})$ remains low when all range-cell detection scores are low. A high segment-level target presence score therefore cannot arise solely from the accumulation of low responses across multiple non-target range cells. 

RID-CGL generates a target pseudo-label at the location with the highest response and applies positive supervision only to the corresponding range cell in each view~\cite{11730983}. In contrast, smooth maximum pooling propagates target presence supervision to multiple range cells, assigning greater supervision weights to those with higher responses. Minimizing $\ell_{\mathrm{pre}}(\mathbf{a})$ can therefore preferentially increase higher detection responses in a target segment while also providing target presence supervision to range cells outside the candidate set. This provides an additional training signal for potential target locations omitted from the candidate set, complementing the location supervision from the refined pseudo-labels.

\subsubsection{Total Loss}
The range-cell loss for a target segment is
\begin{equation}
	\ell_{\mathrm{pos}}=\ell_{\mathrm{fine}}+\ell_{\mathrm{pre}}.
\end{equation}
Let $\mathbb{E}_+$ and $\mathbb{E}_-$ denote the sample means over target and background segments, respectively. To balance the contributions of the two classes during training, we define the total loss as an equally weighted combination of their respective mean losses:
\begin{equation}
	\begin{gathered}
	\mathcal{L}_{\mathrm{total}}=
	\mathcal{L}_{\mathrm{det}}+
	\mathcal{L}_{\mathrm{cls}},\\
	\mathcal{L}_{\mathrm{det}}\!=\!
	\mathbb{E}_+[\ell_{\mathrm{pos}}]
	\!+\!\mathbb{E}_-[\ell_{\mathrm{neg}}],\ \
	\mathcal{L}_{\mathrm{cls}}\!=\!
	\mathbb{E}_+[\ell_{\mathrm{cls}}]
	\!+\!\mathbb{E}_-[\ell_{\mathrm{cls}}].
	\end{gathered}
\end{equation}

\section{Experiments}\label{sec_experiments}

\subsection{Dataset Description}
To evaluate the effectiveness of the proposed training method using measured sea-clutter data, we use data collected in November 2022 at sea states 2--5 from the Maritime Target Data Sharing Program database established by Naval Aviation University~\cite{jian2023sea}. The measurements were acquired with an SPPR50P X-band solid-state coherent radar deployed in Yantai, Shandong, China. Table~\ref{table_Radar_parameters} lists its operating parameters.

\begin{figure*}[t]
	\centering
	\includegraphics[width=\textwidth]{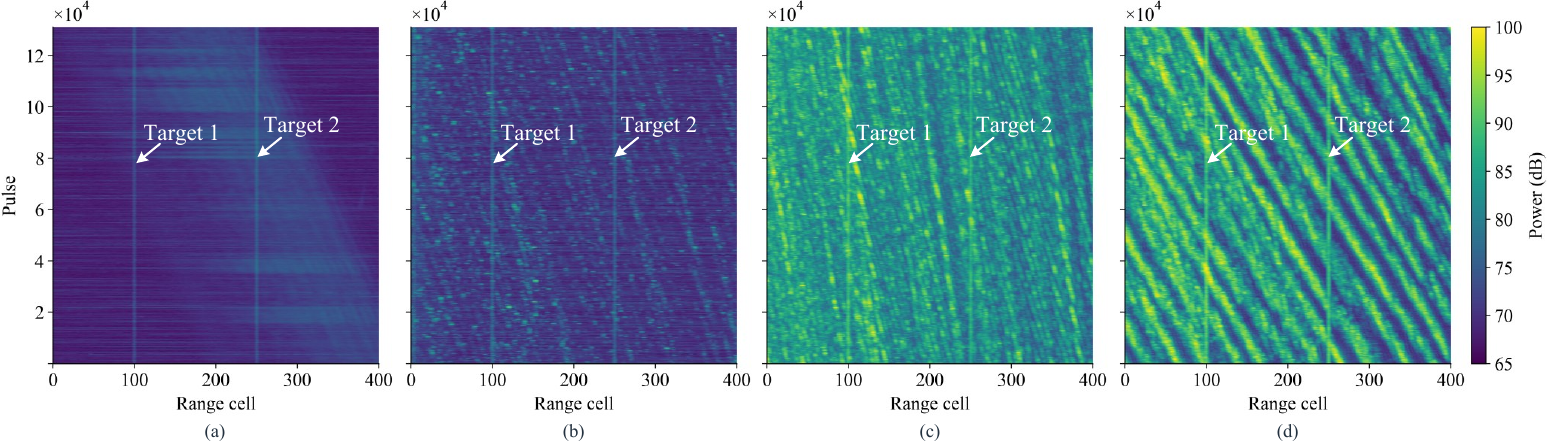}
	\vspace{-0.7cm}
	\caption{Target echoes in sea clutter at an SCR of 2\,dB. (a)--(d) correspond to sea states 2, 3, 4, and 5, respectively. Clutter intensity increases with sea state.}
	\label{fig_four_data_level}
\end{figure*}

\begin{table}[t]
	\renewcommand{\arraystretch}{1.3}
	\caption{Radar Parameters}
	\label{table_Radar_parameters}
	\centering
	\vspace{-0.2cm}
	\setlength{\tabcolsep}{2.0mm}{ 
		\begin{tabular}{cccc}
			\bottomrule
			\cmidrule{1-4} 
			{\textbf{Bandwidth}} & {\textbf{Power}} & {\textbf{Range Resolution}} & {\textbf{Waveform}} \\
			\cmidrule{1-4} 
			25 MHz &  100 W &  6\,m & LFM\\
			\bottomrule
			\cmidrule{1-4}
			{\textbf{Carrier Frequency}} & {\textbf{PRF}} & {\textbf{Sampling Rate}} & {\textbf{Pulse width}} \\
			\cmidrule{1-4} 
			9.4 GHz & 2000 Hz & 60 MHz & 8\,µs \\
			\bottomrule
			\cmidrule{1-4}
	\end{tabular}}
\end{table}

Because the SCR of targets originally present in the dataset cannot be adjusted to prescribed experimental values, we follow the data construction procedure used in our previous work~\cite{11730983} and superimpose simulated target echoes on measured sea clutter. This preserves the measured background while allowing precise control of the target SCR.

Using the signal model in Section~\ref{subsec_Signal_Model} and the parameters in Table~\ref{table_Radar_parameters}, we generate a target echo vector $\mathbf{r}_{\mathrm{tar}}$ centered on range cell $g$. Its peak power before scaling is $P_h(g)=|r_{\mathrm{tar},g}|^2$. To achieve a prescribed SCR of $\rho$\,dB, defined in terms of the target peak power and mean background power, we first estimate the background power from 60 range cells surrounding the target center in the current pulse. We then compute the scaling factor $A_\rho(g)$ and apply it to $\mathbf{r}_{\mathrm{tar}}$ as follows:
\begin{equation}
	\begin{gathered}
		P_c(g)=\frac{1}{60}\sum_{m=g-30}^{g+29}|c_m+n_m|^2,\\
		A_\rho(g)=10^{\rho/20}\sqrt{\frac{P_c(g)}{P_h(g)}},\quad
		\tilde{\mathbf{r}}_{\mathrm{tar}}=A_\rho(g)\mathbf{r}_{\mathrm{tar}}.
	\end{gathered}
	\label{eq_peak_scr_scaling}
\end{equation}
Superimposing the scaled target echo $\tilde{\mathbf{r}}_{\mathrm{tar}}$ on the measured background $\mathbf{c}+\mathbf{n}$ yields the echo vector $\mathbf{r}$ at the specified SCR. Fig.~\ref{fig_four_data_level} shows examples of target echoes added to measured sea clutter under different sea states.

If the target occupies the same range cell in every pulse, a detector may achieve a high $P_{\mathrm d}$ on such echoes simply by producing a high detection statistic in that range cell. The detector may therefore rely on the fixed target location rather than learn discriminative features that distinguish targets from background. To avoid this, we randomly select the target range cell in each training pulse. 

To construct the dataset, we first divide the pulses in each echo data file chronologically into contiguous training, validation, and test sets in a 4:2:4 ratio. Within each set, we then extract nonoverlapping echo segments using a sliding window with both length and stride set to $K=64$ range cells. Each training segment is assigned an incomplete label indicating whether it contains a target.

\subsection{Comparison Methods} \label{section_compar_method}
Because research on training intelligent radar target detectors with incomplete labels remains limited, we select relevant methods from other fields for comparison.

\textbf{SVDD}~\cite{ruff2018deep}: This method trains a feature extractor on normal images for image anomaly detection, mapping image features to a compact region around a hypersphere center. At detection time, it computes an anomaly score based on the distance between the input image's feature vector and the center and uses this score for detection.

\textbf{CVVAE}~\cite{bai2026radar}: This method trains a variational autoencoder on multipulse echoes from each range cell in target-free segments to learn the clutter distribution. At detection time, it constructs a detection statistic based on the discrepancy between the input echo and the autoencoder's reconstruction.

\textbf{LSP}~\cite{wang2019comparison}: This method uses linear softmax pooling to aggregate scores from short audio segments into a recording-level score for sound event detection. It then combines the recording-level score with incomplete labels to construct a loss that guides the model to learn detection scores for individual segments.

\textbf{RID-CGL}~\cite{11730983}: This method, proposed in our previous work, combines segment-level labels, detection responses from a Siamese Transformer, and a translation-equivariance constraint to construct range-cell-level pseudo-labels for detector training.

\textbf{RTFM}~\cite{tian2021weakly}: This method selects several segments with the largest feature magnitudes from each video and trains a video anomaly detector using incomplete labels. To rank segments containing anomalous events higher in terms of feature magnitude, it imposes a feature-magnitude constraint that encourages the selected segments in anomalous videos to have a higher mean feature magnitude than those in normal videos.

\textbf{UMIL}~\cite{lv2023unbiased}: This method divides video segments into a high-confidence group and an ambiguous group according to the stability of the detector's predictions, then clusters the ambiguous segments. It combines classification supervision on the high-confidence group with clustering constraints on the ambiguous group to learn more stable features for anomaly discrimination.

\textbf{FST}~\cite{9375476}: This method uses accurate range-cell-level labels to train a detector with the same architecture and inputs as the proposed detector.

We refer to the preceding six methods trained with incomplete labels as the comparison methods. FST serves as an upper bound on the detection performance of detectors trained using only incomplete labels.

\subsection{Implementation Details} \label{section_Implement_detail}
Unless otherwise specified, we use $K_{\mathrm c}=2$ candidates for label refinement, with respective weights of $0.8$ and $0.2$. Each parallel convolutional branch of the detector has $d_{\mathrm b}=32$ output channels, and the fusion convolution produces multiscale features with $d=64$ channels. The global attention module uses $L=4$ Transformer blocks, each with $N_{\mathrm h}=4$ attention heads of dimension $d_h=16$. The hidden-layer dimension of the MLP is $d_{\mathrm{ff}}=128$.

We use Kaiming initialization for the network parameters and train the detector with the Adam optimizer at an initial learning rate of $10^{-3}$ and a batch size of $4096$ for $300$ epochs. After training, we select the checkpoint with the lowest validation loss for performance evaluation. To reduce the influence of random variations on the experimental results, we repeat each experiment 5 times and report the mean results. The experimental platform is equipped with two Intel Xeon Platinum 8380 CPUs, 1024\,GB of memory, and eight NVIDIA A40 GPUs. We use Ubuntu 22.04.4 LTS as the operating system and PyTorch 2.4.0 as the deep learning framework.

\subsection{Evaluation Metric}
For a fair comparison, we evaluate the detectors trained using different methods in terms of detection probability at a fixed false alarm rate~\cite{10057006,10752981}. For a test set of $N_{\mathrm{test}}$ echo segments, let $\mathcal{K}^{(n)}$ denote the set of target locations in segment $n$. To reduce the influence of target echoes on detection threshold estimation, we exclude the target centers and the surrounding range cells within the guard radius from each segment. The remaining range cells form the background sample set:
\begin{equation}
	\begin{gathered}
		\Omega_0=\{(n,i):n\in\{1,\ldots,N_{\mathrm{test}}\},\\
		i\in\{1,\ldots,K\},\quad |i-k|>h_e,\ \forall k\in\mathcal{K}^{(n)}\}.
	\end{gathered}
\end{equation}
Here, $(n,i)$ denotes the $i$-th range cell in the $n$-th echo segment, and $h_e$ is the guard radius. We sort the detection statistics of the range cells in $\Omega_0$ in descending order as $a_{(1)}\ge\cdots\ge a_{(|\Omega_0|)}$, where the parenthesized subscripts indicate rank. We determine the detection threshold $\gamma$ from the empirical quantile corresponding to the specified false alarm rate $P_{\mathrm{fa}}$. We then apply the decision rule in~\eqref{eq_target_detect} and compute the detection probability as follows:
\begin{equation}
	\gamma=a_{(\lceil|\Omega_0|P_{\mathrm{fa}}\rceil)},\quad
	P_{\mathrm d}=
	\frac{\displaystyle\sum_n\sum_{k\in\mathcal{K}^{(n)}}
		\mathbf{1}[a_k^{(n)}>\gamma]}
	{\displaystyle\sum_n|\mathcal{K}^{(n)}|}.
\end{equation}

\begin{table}[t]
	\renewcommand{\arraystretch}{1.3}
	\caption{Data Files Used for Sea State 5}
	\label{table_5_level_data}
	\centering
	\setlength{\tabcolsep}{3.0mm}
	\vspace{-0.2cm}
	\begin{tabular}{cc}
		\bottomrule
		\cmidrule{1-2}
		{20221112180016\_stare\_HH.mat} & {20221112190020\_stare\_HH.mat} \\
		{20221112205109\_stare\_HH.mat} & {20221112220105\_stare\_HH.mat} \\
		{20221113030022\_stare\_HH.mat} & {20221113060045\_stare\_HH.mat} \\
		\bottomrule
		\cmidrule{1-2}
	\end{tabular}
\end{table}

\begin{figure*}[t]
	\centering
	\includegraphics[width=\textwidth]{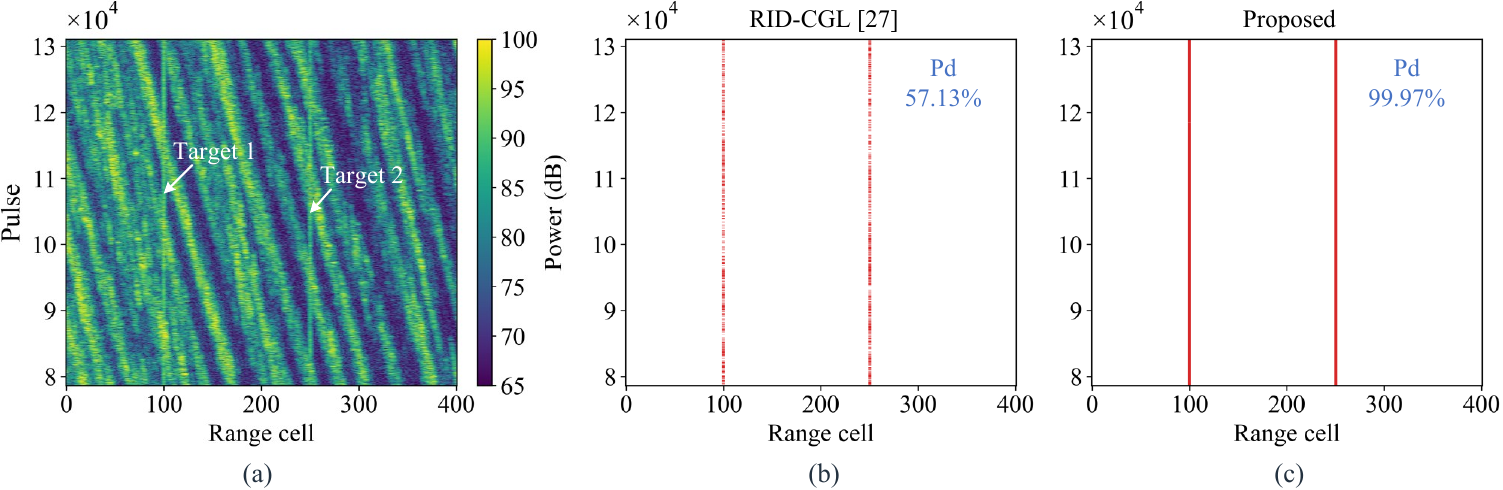}
	\vspace{-0.7cm}
	\caption{Detection results for sea state 5 with two targets, each at an SCR of $2$\,dB. (a) Test echoes from the last 40\% of the pulses in 20221112205109\_stare\_HH.mat. (b) and (c) Detection results at $P_{\mathrm{fa}}=10^{-5}$ for detectors trained with RID-CGL and the proposed method, respectively.}
	\label{fig_detection_result}
\end{figure*}

\subsection{Performance Comparison}
\subsubsection{Visual Comparison}
We divided the pulses in the six sea state 5 echo data files listed in Table~\ref{table_5_level_data} into training, validation, and test sets. We superimposed simulated targets with an SCR of $2$\,dB on the training echoes and used the resulting dataset to train detectors with our previous method, RID-CGL, and the proposed method. To visually compare their detection results, we superimposed two simulated targets at fixed locations, each with an SCR of $2$\,dB, on the test pulses from 20221112205109\_stare\_HH.mat. Fig.~\ref{fig_detection_result} shows the echo power distribution after the targets were added and the detection results of the two detectors. At $P_{\mathrm{fa}}=10^{-5}$, the detector trained with RID-CGL produces intermittent detections at the two target locations, with $P_{\mathrm d}=57.13\%$. The detector trained with the proposed method produces nearly continuous detections at both locations across pulses, with $P_{\mathrm d}=99.97\%$. This result shows that the proposed method substantially improves detector performance when training with incomplete labels.

\begin{figure*}[t]
	\centering
	\includegraphics[width=\textwidth]{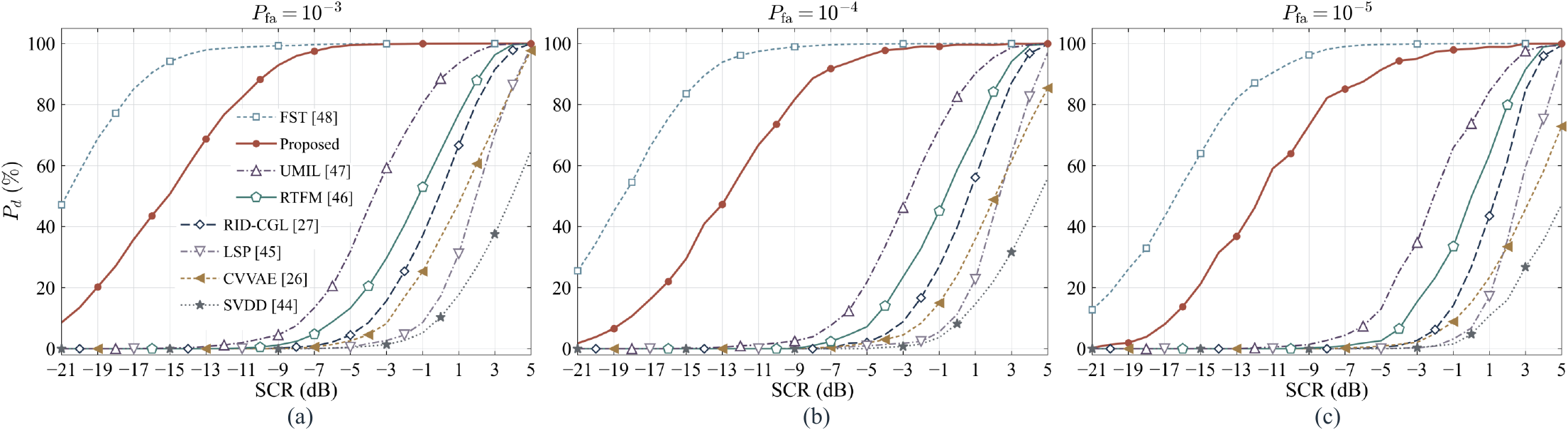}
	\vspace{-0.7cm}
	\caption{$P_{\mathrm d}$ versus SCR for each method at sea state 5. Panels (a), (b), and (c) correspond to $P_{\mathrm{fa}}$ values of $10^{-3}$, $10^{-4}$, and $10^{-5}$, respectively. The legend in (a) applies to all three panels.}
	\label{fig_comparison_curve}
	\vspace{-0.4cm}
\end{figure*}

\subsubsection{Effects of SCR and $P_{\mathrm{fa}}$}
We use the sea state 5 data in Table~\ref{table_5_level_data} to further examine detection performance over a wider range of target strengths and at different $P_{\mathrm{fa}}$ levels. These experiments vary the SCR from $-21$ to $5$\,dB in steps of $1$\,dB and broaden the comparison from RID-CGL to all the comparison methods defined in Section~\ref{section_compar_method}. Fig.~\ref{fig_comparison_curve} shows $P_{\mathrm d}$ as a function of SCR for each method. At each $P_{\mathrm{fa}}$ considered, the $P_{\mathrm d}$ curve for the detector trained with the proposed method lies to the left of those for the competing methods, indicating that the proposed method achieves the best detection performance among them by attaining a given $P_{\mathrm d}$ at a lower SCR. As $P_{\mathrm{fa}}$ decreases from $10^{-3}$ to $10^{-5}$, the curve for the proposed method moves progressively closer to that of the upper-bound reference FST. These results indicate that the detector trained with the proposed method maintains strong detection performance under stringent $P_{\mathrm{fa}}$ constraints.

At $P_{\mathrm{fa}}=10^{-3}$, CVVAE achieves a higher $P_{\mathrm d}$ than LSP. As $P_{\mathrm{fa}}$ decreases, however, $P_{\mathrm d}$ becomes greater for LSP than for CVVAE at higher SCRs. This difference is most pronounced at $P_{\mathrm{fa}}=10^{-5}$. The training objectives of the two methods offer one explanation for this change. CVVAE learns to reconstruct the background using only target-free echoes and uses the reconstruction error as its detection statistic. A lower prescribed $P_{\mathrm{fa}}$ places the detection threshold at a higher quantile of the background reconstruction error distribution. The background reconstruction loss does not directly optimize the probability that the reconstruction error in a target range cell exceeds this threshold~\cite{koizumi2019unsupervised}. In contrast, LSP pools the detection scores of individual range cells into a segment-level score. This allows LSP to use segment-level labels to optimize the pooled score directly. We therefore adopt this discriminative training approach in our proposed method.

\begin{table}[t]
	\renewcommand{\arraystretch}{1.3}
	\centering
	\caption{Mean SCR gain (dB) of the proposed method relative to the other methods at sea state 5.}
	\label{tab_average_db_gap}
	\vspace{-0.2cm}
	\setlength{\tabcolsep}{2.2mm}
	\begin{tabular}{l|ccc|c}
		\bottomrule
		\cmidrule{1-5}
		Method & $P_{\mathrm{fa}}\!=\!10^{-3}$ & $P_{\mathrm{fa}}\!=\!10^{-4}$ & $P_{\mathrm{fa}}\!=\!10^{-5}$ & Avg \\
		\bottomrule
		\cmidrule{1-5}
		FST~\cite{11029508}     & -6.03 & -5.79 & -5.15 & -5.66 \\
		UMIL~\cite{lv2023unbiased}    & 11.24 &  9.72 &  9.45 & 10.14 \\
		RTFM~\cite{tian2021weakly}    & 13.40 & 11.61 & 10.98 & 11.99 \\
		RID-CGL~\cite{11730983} & 14.51 & 12.80 & 12.80 & 13.37 \\
		LSP~\cite{wang2019comparison}     & 16.89 & 15.03 & 14.59 & 15.50 \\
		CVVAE~\cite{bai2026radar}   & 16.06 & 15.07 & 15.26 & 15.46 \\
		SVDD~\cite{ruff2018deep}    & 19.72 & 17.90 & 17.47 & 18.36 \\
		\bottomrule
		\cmidrule{1-5}
	\end{tabular}
\end{table}

\subsubsection{Performance Gain}
To quantify the SCR gain of the proposed method, we calculate the mean difference between the SCRs required by a comparison method and the proposed method to achieve the same $P_{\mathrm d}$ at a given $P_{\mathrm{fa}}$. We further extend the tested SCR range so that, for each method, $P_{\mathrm d}$ approaches 0\% and 100\% at the lower and upper ends, respectively. We then select 21 $P_{\mathrm d}$ levels from 0\% to 100\% in increments of 5 percentage points. For each specified $P_{\mathrm{fa}}$, we use linear interpolation to estimate the SCR required by each method at each $P_{\mathrm d}$ level covered by both curves. The SCR differences are then averaged across these levels. 

Table~\ref{tab_average_db_gap} reports the mean SCR gains at three values of $P_{\mathrm{fa}}$, together with their equally weighted mean in the last column. At each $P_{\mathrm{fa}}$, the proposed method achieves an SCR gain exceeding $9$\,dB over every comparison method. Based on the means in the last column, the gains over RID-CGL and UMIL, the best-performing comparison method on average, are $13.37$ and $10.14$\,dB, respectively. The mean SCR gap between the proposed method and FST is approximately $5$--$6$\,dB at each $P_{\mathrm{fa}}$. This gap narrows as $P_{\mathrm{fa}}$ decreases, indicating that the detection performance of the proposed method approaches that of FST under more stringent false alarm constraints.

\begin{figure}[t]
	\centering
	\includegraphics[width=\columnwidth]{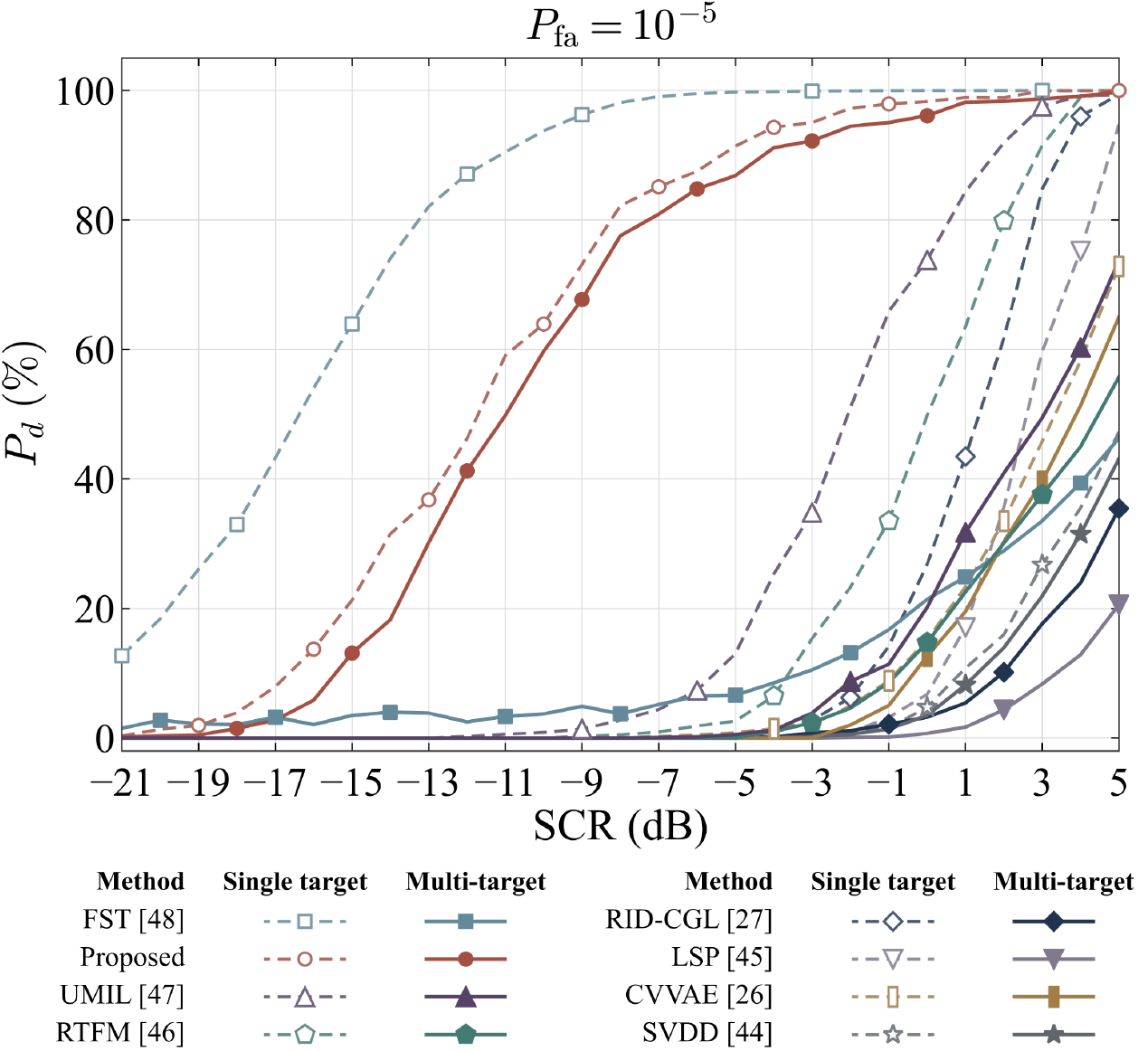}
	\caption{$P_{\mathrm d}$ versus SCR for each method at sea state 5. Dashed and solid curves of the same color give the single-target and multitarget test results, respectively, for the same method.}
	\label{fig_multi_target}
\end{figure}

\subsubsection{Multitarget Detection}
To assess the effect of increasing the number of targets per echo segment on $P_{\mathrm d}$, we compare all methods under single-target and multitarget conditions. Following the test setup in~\cite{10058041}, we construct the multitarget test set from the test pulses in the sea state~5 echo data files listed in Table~\ref{table_5_level_data}. We increase the number of simulated targets superimposed on each target-containing segment from one to two and keep all other test conditions fixed. Fig.~\ref{fig_multi_target} shows $P_{\mathrm d}$ versus SCR for each method in both tests at $P_{\mathrm{fa}}=10^{-5}$. In the multitarget test, the $P_{\mathrm d}$ curve for the proposed method generally lies above those of the other methods. At matched SCR values, the decrease in $P_{\mathrm d}$ from single-target to multitarget testing is small for the proposed method and generally larger for the other methods.

To quantify the loss in detection performance from single-target to multitarget testing, we use the $P_{\mathrm d}$ levels and interpolation procedure described above. At each $P_{\mathrm d}$ level covered by both curves for a given method, we subtract the SCR required in the single-target test from that required in the multitarget test. Table~\ref{table_mul_gap} reports the mean of these differences for each method. The proposed method has a mean SCR loss of only $1.05$\,dB, compared with $3.59$\,dB for RID-CGL, and achieves the smallest loss among the discriminative methods trained with segment-level labels. FST exhibits the largest mean SCR loss, suggesting that discriminative detectors can readily develop a single-target bias when the original training segments contain at most one target each. CVVAE and SVDD have smaller mean SCR losses than the proposed method, at $0.58$\,dB and $0.37$\,dB, respectively. Because both methods are trained exclusively on target-free echoes, they do not rely on target-count cues and are therefore less sensitive to the single-target training bias. However, the lack of supervision from target-containing segments also explains why their $P_{\mathrm d}$ values are generally much lower than those of the proposed method in the multitarget test.

\begin{table}[t]
	\renewcommand{\arraystretch}{1.3}
	\caption{Multitarget performance loss (dB) for each method.}
	\label{table_mul_gap}
	\centering
	\vspace{-0.2cm}
	\setlength{\tabcolsep}{3.5mm}{ 
		\begin{tabular}{cccc}
			\bottomrule
			\cmidrule{1-4} 
			{\textbf{FST}~\cite{11029508}} & {\textbf{Proposed}} & {\textbf{UMIL}~\cite{lv2023unbiased}} & {\textbf{RTFM}~\cite{tian2021weakly}} \\ 
			\cmidrule{1-4} 
			20.5071 &  1.0523 & 4.7327 & 3.7737 \\
			\bottomrule
			\cmidrule{1-4}
			{\textbf{RID-CGL}~\cite{11730983}} & {\textbf{LSP}~\cite{wang2019comparison}} & {\textbf{CVVAE}~\cite{bai2026radar}} & {\textbf{SVDD}~\cite{ruff2018deep}} \\
			\cmidrule{1-4} 
			3.5931 & 3.0360 & 0.5847 & 0.3670 \\
			\bottomrule
			\cmidrule{1-4}
	\end{tabular}}
\end{table}

\begin{figure*}[t]
	\centering
	\includegraphics[width=\textwidth]{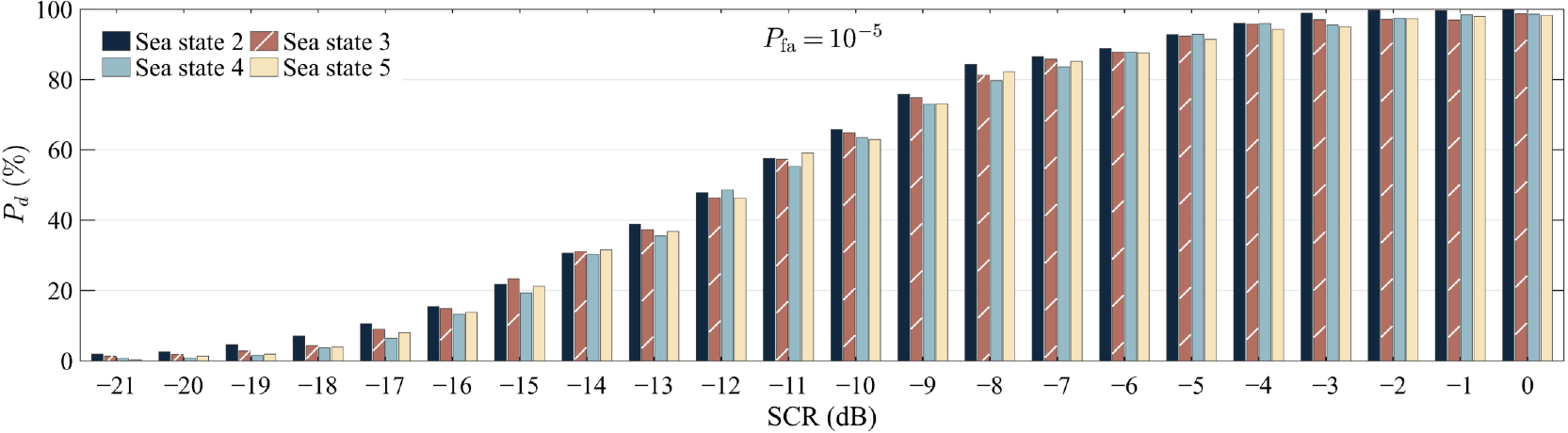}
	\vspace{-0.7cm}
	\caption{$P_{\mathrm d}$ versus SCR at different sea states for the detector trained with the proposed method.}
	\label{fig_diff_sea_state}
	\vspace{-0.4cm}
\end{figure*}

\subsubsection{Effect of Sea State}
To examine whether differences in clutter distributions across sea states affect the detection performance of the proposed method, we supplement the sea state 5 data in Table~\ref{table_5_level_data} with the measured echoes from sea states 2, 3, and 4 listed in Table~\ref{table_all_level_data}. We train a separate detector for each sea state and evaluate its $P_{\mathrm d}$ on the corresponding test set. Fig.~\ref{fig_diff_sea_state} shows $P_{\mathrm d}$ versus SCR for the four sea states at $P_{\mathrm{fa}}=10^{-5}$. At a fixed SCR, the proposed method achieves similar $P_{\mathrm d}$ values across the four sea states, indicating that differences in clutter distributions have little effect on its detection performance under the evaluated conditions. By contrast, detection performance varies more substantially with SCR than with sea state.

\begin{table}[t]
	\renewcommand{\arraystretch}{1.3}
	\caption{Data Files Used at Different Sea States}
	\label{table_all_level_data}
	\centering
	\setlength{\tabcolsep}{1.2mm}
	\vspace{-0.2cm}
	\begin{tabular}{ccc}
		\bottomrule
		\cmidrule{1-3}
		{\textbf{State}} & \multicolumn{2}{c}{\textbf{Data file}} \\
		\bottomrule
		\cmidrule{1-3}
		\multirow{2}{*}{2} & {20221115040022\_stare\_HH.mat} & {20221115050027\_stare\_HH.mat} \\
		& {20221115070036\_stare\_HH.mat} & {20221115080041\_stare\_HH.mat} \\
		\cmidrule{1-3}
		\multirow{3}{*}{3} & {20221114040018\_stare\_HH.mat} & {20221114130044\_stare\_HH.mat} \\
		& {20221114180042\_stare\_HH.mat} & {20221114010004\_stare\_HH.mat} \\
		& {20221114070031\_stare\_HH.mat} & {20221114190046\_stare\_HH.mat} \\
		\cmidrule{1-3}
		\multirow{3}{*}{4} & {20221112140039\_stare\_HH.mat} & {20221112150043\_stare\_HH.mat} \\
		& {20221113090050\_stare\_HH.mat} & {20221113100031\_stare\_HH.mat} \\
		& {20221113170002\_stare\_HH.mat} & {20221114000105\_stare\_HH.mat} \\
		\bottomrule
		\cmidrule{1-3}
	\end{tabular}
\end{table}

\subsection{Ablation Studies}	
\subsubsection{Decoupling Pseudo-Label Generation From Detector Training}
Our previous method, RID-CGL, generates location pseudo-labels from the detector's own predictions and then uses these pseudo-labels to update the detector. Errors in location supervision can therefore accumulate over training iterations. To avoid this coupling, the proposed method uses a background model to generate location pseudo-labels independently of detector training. To determine whether the performance gain comes from this decoupling or from the background model's own detection capability, we apply the background model directly to target detection and compare it with detectors trained using the proposed method and RID-CGL. Table~\ref{tab_background_model} gives the $P_{\mathrm d}$ values of these methods at sea state 5. Across all tested conditions, the background model has a lower $P_{\mathrm d}$ than RID-CGL. Its highest $P_{\mathrm d}$ is only $1.65\%$, whereas the lowest $P_{\mathrm d}$ of the detector trained using the background-model-generated location pseudo-labels is $91.39\%$. These results suggest that the performance gain stems from decoupling pseudo-label generation from detector training rather than from the background model's own detection capability.

To further quantify the contribution of pseudo-label generation decoupling to the improvement in $P_{\mathrm d}$, we compare the proposed method with a variant using detector-generated pseudo-labels (DGPL). DGPL differs from the proposed method only in the source of location pseudo-labels, using the detector's own predictions instead of the background model. All other experimental settings remain unchanged. At the same SCR and $P_{\mathrm{fa}}$, the proposed method consistently achieves a higher $P_{\mathrm d}$ than DGPL, in which pseudo-label generation remains coupled to detector training. This indicates that decoupling location pseudo-label generation from detector training accounts for most of the performance gain over RID-CGL. Moreover, although both DGPL and RID-CGL generate location pseudo-labels from the detector's own predictions, DGPL consistently achieves a higher $P_{\mathrm d}$ than RID-CGL. This difference indicates that the other improvements incorporated into the proposed method also contribute to detection performance.

\begin{table}[t]
	\renewcommand{\arraystretch}{1.3}
	\centering
	\caption{$P_{\mathrm d}$ (\%) for each method at sea state 5}
	\label{tab_background_model}
	\vspace{-0.2cm}
	\setlength{\tabcolsep}{1.6mm}
	\begin{tabular}{c|cc|cc|cc|cc}
		\toprule
		\cmidrule{1-9}
		\multirow{2}{*}{SCR}
		& \multicolumn{2}{c|}{RID-CGL~\cite{11730983}}
		& \multicolumn{2}{c|}{Background}
		& \multicolumn{2}{c|}{DGPL}
		& \multicolumn{2}{c}{Proposed} \\
		& $10^{-4}$ & $10^{-5}$
		& $10^{-4}$ & $10^{-5}$
		& $10^{-4}$ & $10^{-5}$
		& $10^{-4}$ & $10^{-5}$ \\
		\midrule
		\cmidrule{1-9}
		-5 &  2.04 &  0.40 & 0.14 & 0.00 & 13.34 &  6.83 & 95.98 & 91.39 \\
		-4 &  4.67 &  1.03 & 0.31 & 0.01 & 20.27 & 11.26 & 97.80 & 94.31 \\
		-3 &  8.77 &  2.74 & 0.59 & 0.02 & 30.75 & 20.27 & 98.24 & 95.04 \\
		-2 & 16.66 &  6.23 & 0.92 & 0.05 & 42.73 & 30.70 & 99.08 & 97.25 \\
		-1 & 26.99 & 14.16 & 1.22 & 0.11 & 56.97 & 43.66 & 99.04 & 97.92 \\
		0 & 39.85 & 26.82 & 1.47 & 0.17 & 70.77 & 58.43 & 99.62 & 98.29 \\
		1 & 56.12 & 43.47 & 1.65 & 0.22 & 82.35 & 73.45 & 99.64 & 98.89 \\
		\bottomrule
		\cmidrule{1-9}
	\end{tabular}
\end{table}

\subsubsection{Background Statistical Features}
\begin{figure}[t]
	\centering
	\includegraphics[width=\columnwidth]{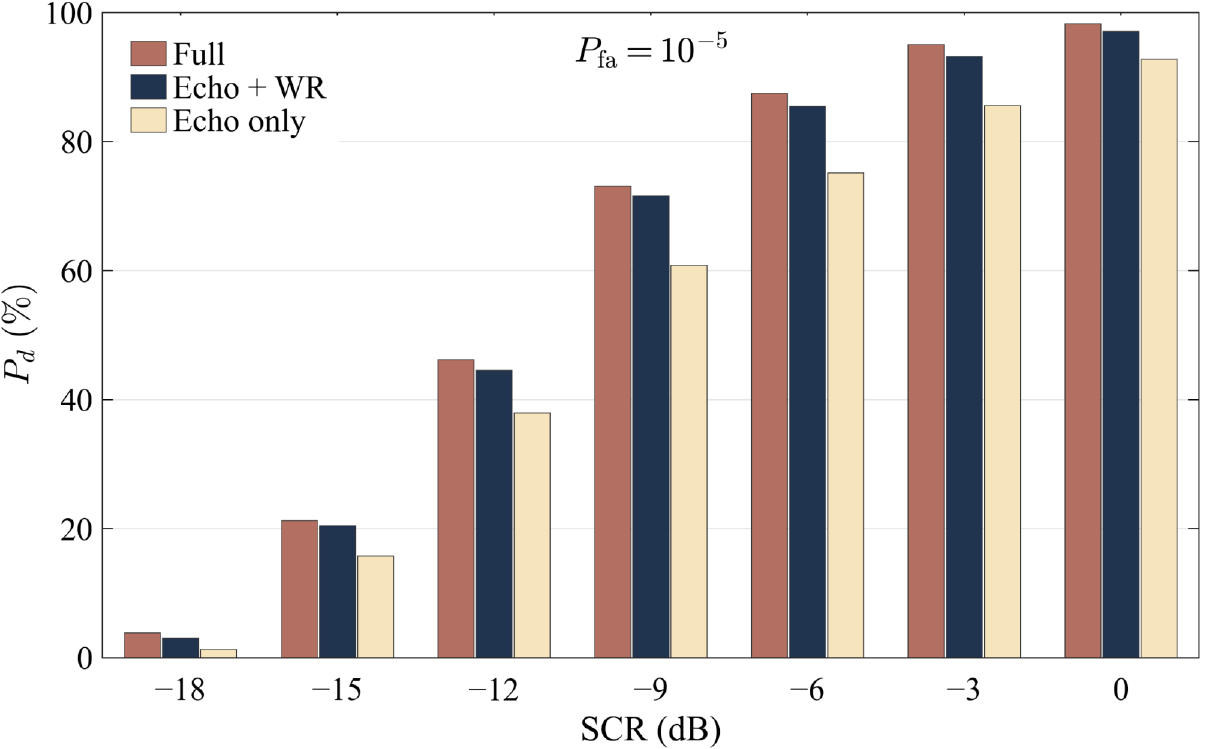}
	\vspace{-0.6cm}
	\caption{$P_{\mathrm d}$ for different input feature combinations. ``Full'' uses the normalized echo, whitened residuals, and log anomaly scores. ``Echo + WR'' omits the log anomaly scores, and ``Echo only'' also omits the whitened residuals.}
	\label{fig_input_ablation}
\end{figure}

In addition to generating location pseudo-labels, the background model provides the detector with two input features, namely the whitened residuals and the log anomaly scores. To assess their contributions to $P_{\mathrm d}$, we train two detector variants using the proposed method, varying only the input feature combinations. We use ``Echo + WR'' to denote the variant retaining the normalized echo and whitened residuals (WR) and ``Echo only'' to denote the variant retaining only the normalized echo.
The complete detector, denoted as ``Full'', uses the normalized echo, whitened residuals, and log anomaly scores. Fig.~\ref{fig_input_ablation} shows the $P_{\mathrm d}$ values of the three detectors at $P_{\mathrm{fa}}=10^{-5}$. Removing the log anomaly scores reduces $P_{\mathrm d}$ relative to ``Full''.
Further removing the whitened residuals leads to an additional reduction in $P_{\mathrm d}$. At every SCR tested, the reduction from removing the whitened residuals is larger than that from removing the log anomaly scores. This ablation study shows that both background statistical features improve $P_{\mathrm d}$, with a greater contribution from the whitened residuals.

\subsubsection{Multiscale Features}
\begin{figure}[t]
	\centering
	\includegraphics[width=\columnwidth]{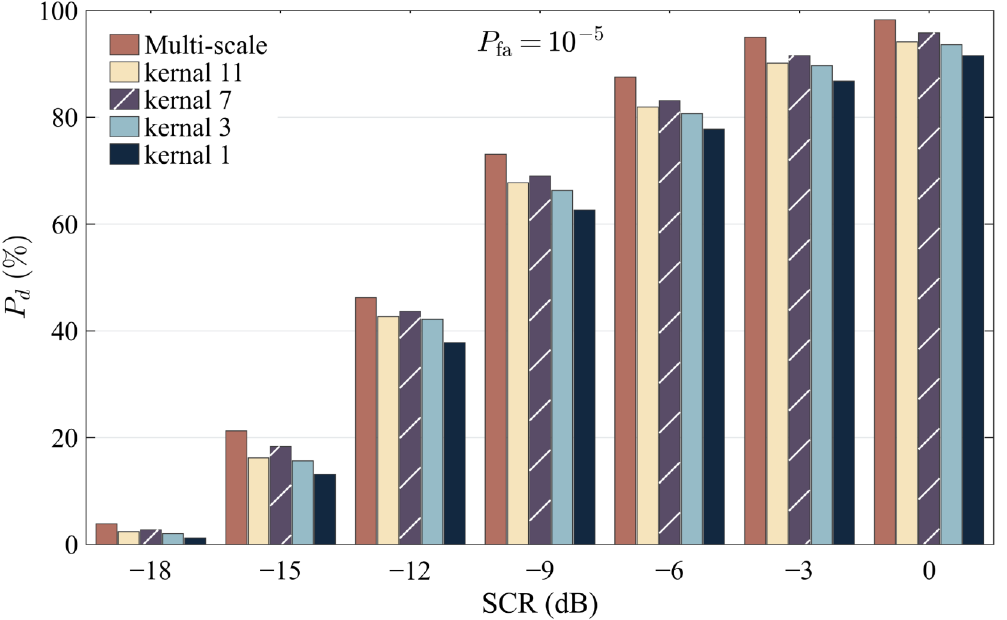}
	\vspace{-0.6cm}
	\caption{Ablation of multiscale features. The proposed detector, labeled Multi-scale, extracts local features in parallel using convolution kernels of widths $1$, $3$, $7$, and $11$. Each of the four single-scale variants uses one of these widths.}
	\label{fig_multi_scale_ablation}
\end{figure}

To examine whether combining local features at different scales improves $P_{\mathrm d}$, we compare the proposed background-aware detector with four single-scale variants. The proposed detector extracts local features in parallel using convolutional kernels of widths $1$, $3$, $7$, and $11$. Each single-scale variant uses kernels of only one of these widths. Fig.~\ref{fig_multi_scale_ablation} shows the $P_{\mathrm d}$ values of the detectors at $P_{\mathrm{fa}}=10^{-5}$. 

At every SCR tested, the proposed detector achieves a higher $P_{\mathrm d}$ than all four single-scale variants. The variant with kernel width $1$ has the lowest $P_{\mathrm d}$ because its local convolution combines features only across channels within the current range cell and cannot directly capture echo structure across adjacent range cells. $P_{\mathrm d}$ increases as the kernel width increases from $1$ to $3$ and then to $7$, but decreases when the width increases further to $11$. This decrease occurs because the wider kernel captures clutter variations over a larger neighborhood at the expense of fine details in the target echo across adjacent range cells. These results show that increasing kernel width alone does not consistently improve $P_{\mathrm d}$. By extracting local features with kernels of different widths in parallel and fusing the resulting features, the proposed detector captures both local target echo details and clutter variations over a wider neighborhood, achieving the highest $P_{\mathrm d}$ among the compared detectors.

\subsubsection{Multiple Candidate Range Cells}
\begin{figure}[t]
	\centering
	\includegraphics[width=\columnwidth]{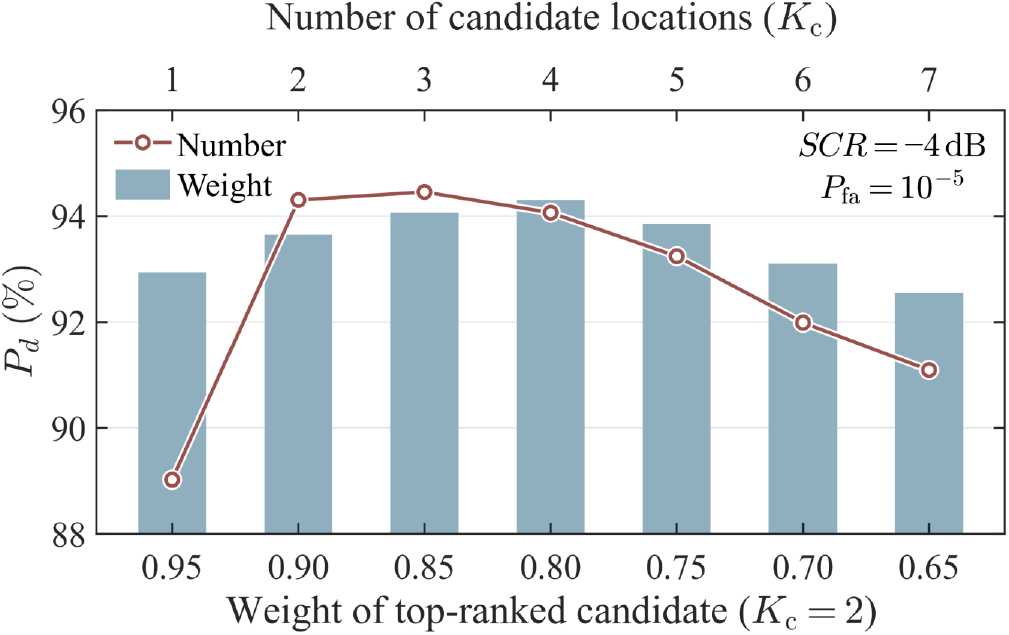}
	\vspace{-0.6cm}
	\caption{Effects of candidate count and candidate weights on $P_{\mathrm d}$. The line uses the upper horizontal axis and shows $P_{\mathrm d}$ as the number of candidates $K_{\mathrm c}$ varies. The bars use the lower horizontal axis and show $P_{\mathrm d}$ as the weight assigned to the higher-scoring range cell varies, with $K_{\mathrm c}$ fixed at 2.}
	\label{fig_distance_candidates}
\end{figure}

When we generate refined pseudo-labels for target-containing segments, the true target range cell may be mislabeled as background if its anomaly score is not the highest within the segment. To reduce this risk, we select the $K_{\mathrm c}$ highest-scoring range cells as candidate locations for the same target and assign weights according to their score rankings. We conduct two ablation experiments to assess the effects of candidate number and weight allocation on $P_{\mathrm d}$. In the first experiment, we increase the number of candidates $K_{\mathrm c}$ from 1 to 7 in increments of one. In the second experiment, we fix $K_{\mathrm c}=2$ and examine the sensitivity of $P_{\mathrm d}$ to weight allocation by progressively decreasing the weight assigned to the higher-scoring range cell from $0.95$ to $0.65$. Fig.~\ref{fig_distance_candidates} shows the experimental results at $SCR=-4\,\mathrm{dB}$ and $P_{\mathrm{fa}}=10^{-5}$.

Increasing the number of candidates from 1 to 2 raises $P_{\mathrm d}$ from approximately 89\% to approximately 94\%. This gain can be explained by how the second-highest-scoring range cell is labeled. When the true target range cell has the second-highest anomaly score, it receives positive supervision with two candidates but is mislabeled as background when only the highest-scoring range cell is selected. Adding a third candidate yields only a small further gain, and $P_{\mathrm d}$ then decreases gradually as the number of candidates increases from 3 to 7. This pattern reflects the competing effects of including lower-ranked range cells. A larger candidate set can include a true target range cell that would otherwise be excluded from positive supervision, but it also assigns positive supervision to more background range cells, potentially interfering with the detector's learning of target locations.

With $K_{\mathrm c}=2$, $P_{\mathrm d}$ first increases and then decreases as the weight assigned to the highest-scoring range cell is reduced. When too much weight is assigned to the top-ranked candidate, location supervision relies primarily on a single anomaly peak and is sensitive to errors in candidate ranking. A moderate increase in the weight of the other candidate reduces this dependence and allows both candidates to provide location cues. As more weight shifts to the other candidate, however, the influence of the location information conveyed by the anomaly score ranking gradually diminishes. The less reliable candidate then has a greater influence on the aggregated score and supervision gradients, which may hinder the learning of discriminative features for distinguishing targets from background. The candidate weights must therefore balance the risk of mislabeling caused by excessive reliance on a single candidate against the need to preserve the location information conveyed by the anomaly score ranking.

\begin{figure}[t]
	\centering
	\includegraphics[width=\columnwidth]{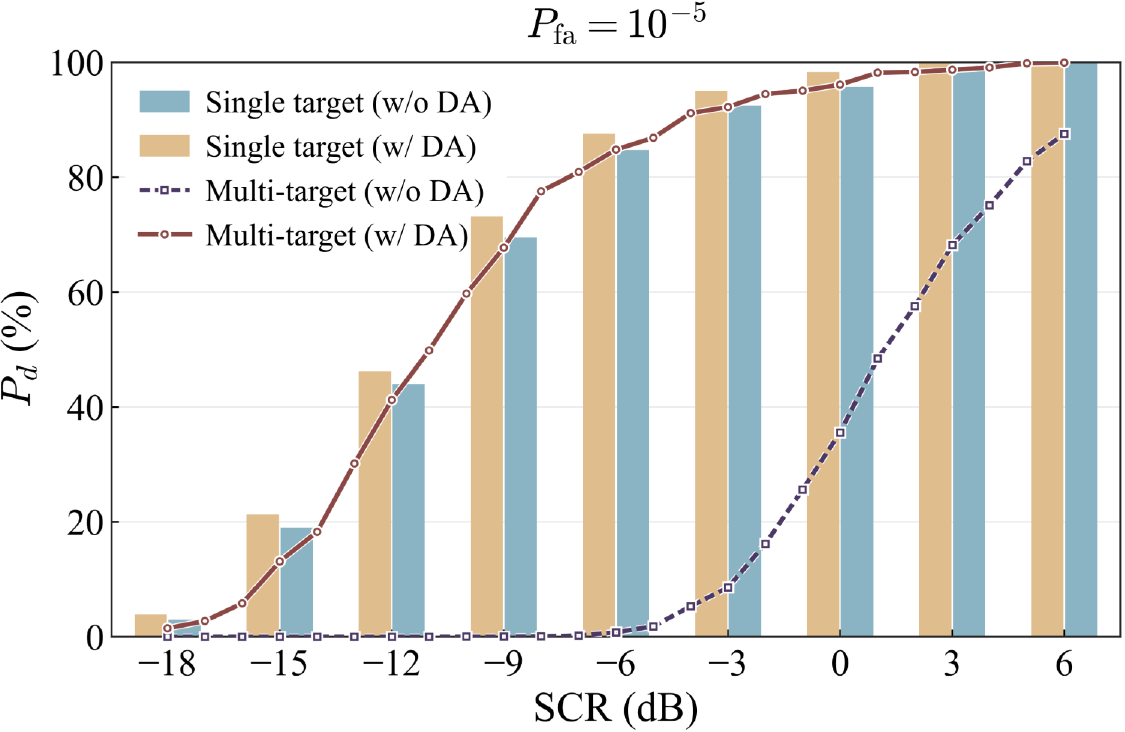}
	\vspace{-0.6cm}
	\caption{Effect of the proposed echo augmentation method on $P_{\mathrm d}$ in single-target and two-target tests. Bars show the single-target results, and lines show the two-target results. In the legend, w/ DA and w/o DA denote training with and without echo augmentation, respectively.}
	\label{fig_data_augmentation}
\end{figure}

\subsubsection{Data Augmentation}
The proposed echo augmentation method superimposes two target-containing echo segments to generate multitarget training samples and improve multitarget detection performance. It also superimposes two clutter-only segments to create more complex clutter backgrounds and combines a clutter-only segment with a single-target echo segment to generate single-target samples in which the target is harder to distinguish from the background. Training on the resulting clutter-only and single-target samples helps the detector learn to distinguish target echoes from clutter fluctuations in more complex backgrounds, thereby contributing to improved single-target detection performance.

To evaluate whether echo augmentation improves single-target and multitarget detection, we train detectors with and without echo augmentation and compare their $P_{\mathrm d}$ values on single-target and two-target test sets at the same SCR and $P_{\mathrm{fa}}$. Fig.~\ref{fig_data_augmentation} shows $P_{\mathrm d}$ versus SCR for both configurations. The results show that constructing training samples by superimposing different types of echo segments substantially improves $P_{\mathrm d}$ in the two-target test while also providing a smaller improvement in the single-target test.

\begin{figure}[t]
	\centering
	\includegraphics[width=\columnwidth]{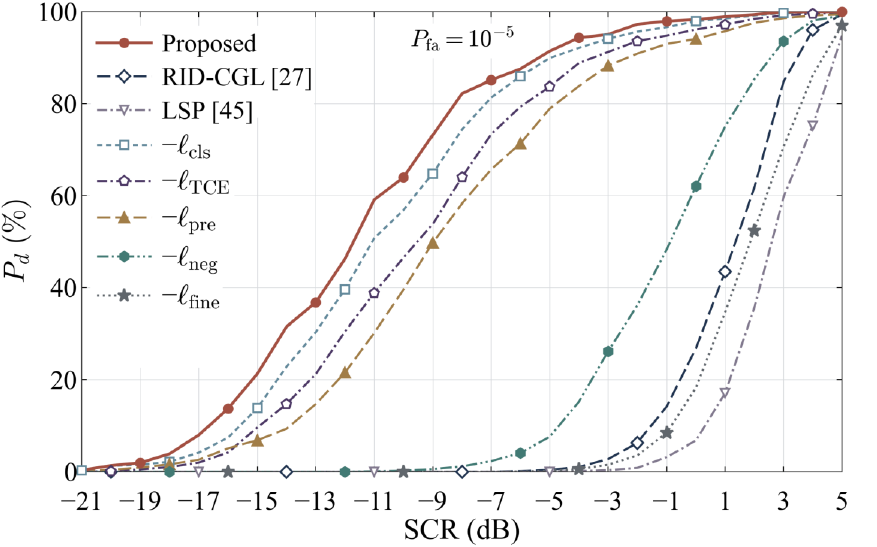}
	\vspace{-0.6cm}
	\caption{Loss function ablation. $-\ell_{\mathrm{cls}}$, $-\ell_{\mathrm{pre}}$, $-\ell_{\mathrm{neg}}$, and $-\ell_{\mathrm{fine}}$ denote removal of the auxiliary classification, segment-level presence, background segment, and refined pseudo-label losses, respectively, one at a time. $-\ell_{\mathrm{TCE}}$ denotes replacement of fourth-order TCE by BCE in the refined pseudo-label loss.}
	\label{fig_loss_ablation}
\end{figure}

\subsubsection{Loss Functions}
To assess the contribution of each term in the proposed multilevel loss, we construct four variants by individually removing the classification loss, background segment loss, refined pseudo-label loss, or segment-level presence loss from the full loss. In a fifth variant, we replace the fourth-order TCE used for target candidate supervision and suppression outside the candidate set in the refined pseudo-label loss with BCE, while keeping all other settings unchanged. Fig.~\ref{fig_loss_ablation} shows the $P_{\mathrm d}$ curves for the complete method, five ablation variants, and two comparison methods, RID-CGL and LSP.

Removing any one of the four losses shifts the $P_{\mathrm d}$ curve to the right relative to the complete method, indicating that all four losses contribute to improved $P_{\mathrm d}$. The reduction is larger when either the background segment loss or the refined pseudo-label loss is removed. These larger reductions can be attributed to the key roles of these two losses in suppressing clutter responses and localizing targets, respectively. In a target-free segment, every range cell can be reliably labeled as background. Removing the background segment loss eliminates this range-cell-level supervision and weakens the detector's ability to suppress strong clutter responses.  Consequently, the detection statistic at the target range cell is less likely to exceed the threshold corresponding to the specified $P_{\mathrm{fa}}$, which explains the substantial drop in $P_{\mathrm d}$. In target-containing segments, removing the refined pseudo-label loss eliminates direct supervision at candidate target locations. Although segment-level presence supervision remains, high responses from background range cells can also increase the segment-level score, making it difficult to accurately localize targets using this supervision alone. This limitation also helps explain why the variant without the refined pseudo-label loss generally achieves a lower $P_{\mathrm d}$ than RID-CGL, which retains location supervision, as shown in Fig.~\ref{fig_loss_ablation}.

\section{Conclusion}\label{sec_conclusion}
We propose a method for training radar target detectors using only segment-level target presence labels, without requiring accurate range-cell-level annotations. The method establishes a statistical background model from target-free echoes and uses it to generate fine-grained pseudo-labels with multiple weighted candidate target locations for each target-containing segment. This makes location supervision independent of the detector's current predictions and reduces the risk of repeatedly reinforcing prediction errors through pseudo-labels during iterative training. A multilevel loss then combines location cues from these fine-grained pseudo-labels with target presence information from the segment-level labels to jointly supervise detector training. To improve multitarget detection, we construct multitarget training samples by linearly superimposing the complex echoes from two target-containing segments. The detector takes normalized echoes, whitened residuals, and log anomaly scores as inputs so that it can use both echo morphology and background information. Parallel convolutions extract multiscale local features, which are then fused with background information across the entire echo segment through global attention to better distinguish target echoes from clutter fluctuations.

Experiments on publicly available measured radar echoes demonstrate that the proposed method outperforms other incomplete-label training methods and substantially narrows the performance gap relative to a detector trained with accurate range-cell labels. Compared with our previous method, RID-CGL, the proposed method achieves an SCR gain of more than $13\,\mathrm{dB}$. Ablation studies further demonstrate the contributions of the key design choices to improved detection performance. It should be noted, however, that the proposed method is currently designed primarily for point targets. Future work will investigate training methods for extended-target detection under incomplete labeling by incorporating the echo characteristics of extended targets.

\appendices 
{	
	\section{Derivation of the Sampling Probabilities}\label{proof_1}
	Let $p\in(0,1)$ be the proportion of target segments in the training set, and sample the base segment according to the training class proportions. If the base contains a target, we always apply superposition and denote the probability that the donor also contains a target by $\pi_+$. If the base is a background segment, we apply superposition with probability $1/2$ and, when it is applied, draw the donor uniformly from the training set. Let $J$ denote the number of constituent segments that contain a target. When both base and donor contain a target, the example is a multitarget case ($J=2$). When only one contains a target, it is a single-target case ($J=1$). Thus,
	\begin{equation}
		\begin{gathered}
			\Pr(J=2)=p\pi_+,\\
			\Pr(J=1)=p(1-\pi_+)+\frac{1}{2}(1-p)p.
		\end{gathered}
	\end{equation}
	Equating the probabilities of generating these two types of examples, $\Pr(J=1)=\Pr(J=2)$, gives
	\begin{equation}
		\pi_+=\frac{3-p}{4},\qquad 1-\pi_+=\frac{1+p}{4}.
	\end{equation}
	These are the donor probabilities for a target base segment. Together with the $1/2$ probability of augmenting a background base segment, they ensure equal sampling probabilities for single-target and multitarget examples. Substitution gives
	\begin{equation}
		\begin{gathered}
			\Pr(J=1)=\Pr(J=2)=\frac{p(3-p)}{4},\\
			\Pr(J=0)=1-\frac{p(3-p)}{2}.
		\end{gathered}
	\end{equation}
	The probability of obtaining a target-containing example after augmentation is therefore
	\begin{equation}
		\begin{gathered}
			\tilde p=\Pr(J\ge1)=\frac{p(3-p)}{2},\\
			\tilde p-p=\frac{p(1-p)}{2}>0.
		\end{gathered}
	\end{equation}
	Among augmented examples that contain targets, the conditional probabilities of the two cases satisfy
	\begin{equation}
		\Pr(J=1\mid J\ge1)=\Pr(J=2\mid J\ge1)=\frac{1}{2}.
	\end{equation}
	Superposition therefore increases the probability that a training example contains a target. Conditional on target presence, the detector is equally likely to receive a single-target or a multitarget example under this sampling rule.
	
}
\bibliographystyle{IEEEtran}
\bibliography{myref}
\end{document}